# OBSTransformer: A Deep-Learning Seismic Phase Picker for OBS Data Using Automated Labelling and Transfer Learning

**Alireza Niksejel[1], Miao Zhang[1*]**

[1]Department of Earth and Environmental Sciences, Dalhousie University, Halifax, Nova Scotia, Canada.

[*]Corresponding Author: miao.zhang@dal.ca

**Key words:** Machine learning, Transfer learning, Ocean bottom seismometer, Seismic phase picking

## Abstract

Accurate seismic phase detection and onset picking are fundamental to seismological studies. Supervised deep-learning phase pickers have shown promise with excellent performance on land seismic data. Although it may be acceptable to apply them to OBS (Ocean Bottom Seismometers) data that are indispensable for studying ocean regions, they suffer from a significant performance drop. In this study, we develop a generalised transfer-learned OBS phase picker – OBSTransformer, based on automated labelling and transfer learning. First, we compile a comprehensive dataset of catalogued earthquakes recorded by 423 OBSs from 11 temporary deployments worldwide. Through automated processes, we label the P and S phases of these earthquakes by analysing the consistency of at least three arrivals from four widely used machine learning pickers (EqTransformer, PhaseNet, Generalized Phase Detection, and PickNet), as well as the AIC picker. This results in an inclusive OBS dataset containing ~36,000 earthquake samples. Subsequently, we employ this dataset for transfer learning and utilize a well-trained land machine learning model – EqTransformer as our base model. Moreover, we extract 25,000 OBS noise samples from the same OBS networks using the Kurtosis method, which are then used for model training alongside the labelled earthquake samples. Using three groups of test datasets at sub-global, regional, and local scales, we demonstrate that OBSTransformer outperforms EqTransformer. Particularly, the P and S recall rates at large distances (>200 km) are increased by 68% and 76%, respectively. Our extensive tests and comparisons demonstrate that OBSTransformer is less dependent on the detection/picking thresholds and is more robust to noise levels.

# 1. Introduction

Seismic phase picking plays a crucial role in land- and ocean-based seismological studies such as source parameter determination and velocity structure inversion. Seismic phase arrivals are traditionally picked manually or by automatic algorithms, which are based on the attribute differences between earthquake signals and background noises. For example, the amplitude/energy change-based methods such as the short-term average/long-term average algorithm (STA/LTA; Allen 1978), data statistics-based approaches such as the AIC (Akaike information criterion) method (Maeda 1985), Kurtosis and Skewness (Saragiotis et al. 2002), etc. These algorithms have been widely used in seismological studies, but they may be insufficient for accurate earthquake location, particularly when applied to low-magnitude events and/or noisy data.

Recent advances in computer processing capabilities, coupled with the advent of large earthquake datasets, have paved the way for the development of several deep learning (DL) models for earthquake detection and phase picking with excellent accuracy and efficiency on land data (e.g., Ross et al. 2018; Wang et al. 2019; Zhou et al. 2019; Zhu & Beroza 2019; Zhu et al. 2019; Mousavi et al. 2020; Xiao et al. 2021). Those DL pickers have been extensively adopted across numerous seismological applications. For example, Liu et al (2020) adopted the DL phase picker, PhaseNet (Zhu & Beroza 2019), to build a high-precision aftershock catalogue for the 2019 Ridgecrest earthquake sequence, leading to more than double the number of events than the routine catalogue. Tan et al. (2021) leveraged PhaseNet to analyze the fault activation processes during the 2016-2017 seismic sequence in central Italy, resulting in an enhanced catalogue including an order of magnitude more earthquakes than the routine catalogue.

Compared to land data, OBS (Ocean Bottom Seismometer) data have received little attention in the development of DL phase pickers, due mostly to the lack of labelled large datasets. Whereas seventy percent of the Earth's surface is covered by the ocean, the home to various seismically active environments such as subduction zones, mid-ocean ridges, transform faults, etc. OBSs provide indispensable observations on offshore earthquakes and their associated tectonic processes, which were highlighted by several recent experiments, for example the Cascadia Initiative (e.g. Scherwath et al. 2011) and the Alaska Amphibious Community Seismic Experiment (AACSE; Obana et al. 2015). However, phase picking on OBS data is much more challenging because they are exposed to various types of ocean-specific noise, such as wind-driven waves, marine mammal sounds, anthropogenic activities, and water column reverberations (e.g. Webb 1998). Recently, seismologists have directly adopted land DL phase pickers to pick seismic phases on OBS data. For instance, Gong et al. (2022) employed EqTransformer (EqT, Mousavi et al. 2020) to analyse seven months of OBS data in the Quebrada system at the East Pacific Rise and detected numerous earthquakes associated with two deep seismicity clouds. Chen et al. (2022) adopted EqT to pick eight months of OBS data deployed in the Challenger Deep in the Southernmost Mariana Trench and delineated an outer-rise fault penetrating depths of 50 km. However, despite DL phase pickers outperform the traditional picking algorithms, they exhibit significant performance drop when applied to OBS data (e.g. Ruppert et al. 2023), due mostly to the different waveform features between land and OBS data.

Transfer learning is an efficient strategy for model training when labelled data are limited or existing models perform poorly (Lapins et al. 2021). It improves the predictive function of a target task (e.g. picking on OBS data) by transferring the knowledge from a (partly-)connected source task (e.g. picking on land data). Transfer learning has been approved its success in seismic phase picking on land data. For example, Chai et al. (2020) transferred the picking capability of PhaseNet (Zhu & Beroza 2019) from natural earthquakes to hydraulic fracturing-induced earthquakes. The dataset used for transfer learning was ~220 times smaller than that of the base model. Lapins et al. (2021) developed a phase-picking model with different task types by re-training the Generalized Phase Detection (GPD) model (Ross et al. 2018). Zhu et al. (2022) adopted an extensive China dataset (i.e. DiTing; Zhao M et al. 2022) to train PhaseNet from scratch and, subsequently, to customise pickers for different regions using transfer learning. Ni et al. (2023) transfer-learned EqT using reliably located earthquakes and explosions in the ANSS Comprehensive Earthquake Catalogue, resulting in an AI-ready dataset with 2% and 85% more P- and S-arrivals.

In this paper, we build a global OBS training dataset using a fully automatic labelling approach and leverage transfer learning to develop an efficient OBS phase picker, named OBSTransformer (called OBST as abbreviation), using an existing, extensively trained land phase picker- EqT as base model (Fig. 1). We compare the performance of OBST with the original EqT model on three groups of manually labelled OBS test datasets from different sources at different scales. We demonstrate that OBST can significantly boost the picking performance, particularly regarding recall rates (see Section 5 for details).

## 2. Data

### 2.1 OBS Dataset Preparation

We collected open-access earthquake waveforms recorded during 11 temporary OBS deployments worldwide to ensure diverse seismic attributes from different submarine environments (Fig. 2). The three-component waveforms of ~100,000 earthquakes were acquired from the data centre of the Incorporated Research Institutions for Seismology (IRIS) using the ObsPy package (Beyreuther et al. 2010; Krischer et al. 2015). All earthquakes are from the available routine catalogues in IRIS, except for the Gorda plate (3,649 events) and East Pacific Rise (23,592 events) regions, which are available from Guo et al. (2021) and Gong et al. (2022) respectively. We performed the following steps for OBS waveform pre-processing: 1) removing mean and linear trends from waveforms, 2) filtering waveforms in an optimal frequency of 3-20 Hz (see Section 4 for detailed discussions and tests), 3) resampling data at 100 Hz, consistent with the original sampling rate of EqT, 4) cutting 60 s earthquake waveform randomly starting from 2-10 s before the theoretical P arrival time, similar to the data preparation in EqT, 5) discarding waveforms contaminated by signals of un-catalogued earthquakes using the STA/LTA method: as pointed out by Mousavi, Zhu et al. (2019), these unexpected signals would lead to labelling errors by mislabelling the uncatalogued earthquakes as noise or vice versa, thus deteriorate the training performance, 6) labelling P and S phases using an automatic approach (see Section 2.2), 7) excluding waveforms with only P or S labels to avoid the potential class imbalance problem during model training, and 8) discarding low quality earthquake samples based on the Signal-to-Noise Ratios (SNR) of both P and S phases, which are defined as below:

$$SNR = 10 \log_{10} \frac{E_S}{E_N}, \tag{1}$$

where $E_S$ and $E_N$ are the waveform energy in a 4-s window before and after each phase, respectively. We remove all samples with negative SNR for either P or S arrivals to reduce the risk of mislabelling.

## 2.2 Pick Auto-Labelling

Well-labelled earthquake signals are crucial for training a robust and reliable DL phase picker. Biased labelling can result in incorrect pattern recognition by the model and lead to suboptimal performance when applied to unseen data. Although many training datasets are based on manual picking of seismic phases, it is challenging for noisy data as the noise can obscure the phase arrivals. Additionally, manual labelling is time-consuming and can be subjective, leading to variability in pick quality within a dataset. A hybrid approach of seismic arrival labelling has become more popular with the advent of traditional and DL phase pickers. This approach involves automated algorithms to quickly pick the seismic phase arrivals, followed by manual review and refinement by human experts to ensure accuracy and consistency. The Stanford Earthquake Dataset (STEAD, Mousavi, Sheng et al. 2019) is an example of such a dataset, where about 30% of the total ground truth is determined by automatic algorithms in seismic networks or a DL phase picking model. To make sure the reliability of those auto labels, this manual verification process is particularly important but takes considerable time.

DL models with different network structures may learn and extract similar features and patterns relevant to earthquake signals, resulting in consistent picks on a given input. However, the likelihood of a complete agreement among different models depends on several factors, such as the quality and quantity of training data, signal complexity, noise level, etc. In this study, we propose and develop a fully automatic strategy to annotate our training OBS dataset using four popular DL phase pickers – EqT (Mousavi et al. 2020), PhaseNet (Zhu & Beroza 2019), GPD (Ross et al. 2019), and PickNet (Wang et al. 2019), along with the AIC auto picker from the ObsPy package (Maeda 1985; Beyreuther et al. 2010):

1) EqT (Mousavi et al. 2020) is a state-of-the-art DL model for joint earthquake detection and P&S arrival picking at local and regional distances. The model was trained on 1 M earthquake and 300 K noise samples from the STEAD at distances up to 300 km (Mousavi, Sheng et al. 2019). The inputs of EqT are 60 s three-component waveforms at 100 Hz sampling rate, and the outputs are three probability distributions for detection, P arrival, and S arrival. Being stacked of one deep encoder and three independent decoders, the EqT model leverages cutting-edge DL techniques such as convolutional layers (CNNs; LeCun et al. 1995), long short-term memory layers (Hochreiter & Schmidhuber 1997), transformers (Vaswani et al. 2017), and self-attention units (Luong et al. 2015) to enhance earthquake detection and phase picking performance (Mousavi et al. 2020). An attention mechanism directs the focus of the network to the valuable parts of input associated with the earthquake signal. Then a transformer helps to pass the driven earthquake information to the subsequent decoders.

2) PhaseNet (Zhu & Beroza 2019) is a 5-layer-deep U-Net-based earthquake detection and arrival time picking model. PhaseNet architecture consists of four down-sampling and four up-sampling stages, including 1-D convolutions and rectified linear units (ReLU) where skip connections are added to concatenate the output of each down-sampling stage to the corresponding up-sampling stage with the same depth to improve convergence during training. PhaseNet was trained using 779,514 recordings based on the Northern California Earthquake Data Center Catalogue at distances up to ~100 km. It operates on 30 s long three-component waveforms as inputs and predicts the probability distributions of P pick, S pick and noise as output (Zhu & Beroza 2019).

3) GPD (Ross et al. 2019) is a six-layer network of four convolutional layers as feature extractors and two fully-connected layers to perform the classification task. The outputs are the classification of P-wave, S-wave, or noise based on a 4-s window at 100 Hz sampling rate. It was trained on 4.5 M hand-labelled three-component waveforms from the Southern California Seismic Network at distances less than 100 km.

4) PickNet (Wang et al. 2019) architected as a fully convolutional network, a variant of the VGG-16 model (Simonyan & Zisserman 2014), plus Rich Side-output Residual Networks (C. Liu et al. 2017) combined with residual units to extract multiscale convolutional information at different levels. The model was trained using data from Hi-net network including ~460,000 P- and ~280,000 S-wave samples recorded at epicentral distances up to ~1,000 km. PickNet includes two independent models for P and S arrival-time picking: one uses only 12-s-long windows from the vertical component, and the other performs on 16-s-long windows from the radial and transverse components of the seismograms as its inputs.

5) AIC (Maeda,1985) is a model selection technique traditionally used for picking the onset of seismic phases. The algorithm assumes that nonstationary distributions (i.e. microseismic signals) can be estimated by locally stationary segments where each can be treated as an autoregressive process (Sleeman & Van Eck 1999). Maeda (1985) retrieved the AIC value for phase picking from the time series based on the following equation:

$$AIC(k) = k \cdot \log var\{x(1,k)\} + (N - K - 1) \log(var\{x(k+1, N)\}), \quad (2)$$

where k, N, and var{x} denote each time series sample, the total number of the input samples, and the variance function, respectively.

    We apply the selected DL phase pickers for phase picking via the SeisBench library (Woollam et al. 2022), except for PickNet, which is unavailable in the library. SeisBench is a novel open-source framework explicitly developed for deploying DL algorithms in seismology. The framework is designed to provide a standardised interface for accessing DL models and datasets while offering a range of processing and data augmentation operations through the API. EqT, PhaseNet and GPD run on continuous waveforms for phase picking. Reference picks are required for the AIC and PickNet pickers, which are referred to the picks by EqT. To ensure high-quality labels, we experimentally set the P and S picking probability thresholds of EqT, PhaseNet, and GPD to 0.5, 0.5, and 0.95. For PickNet, the default picking threshold was adopted. The choices of input waveform windows are followed by each software requirements. In applying AIC, we consider a 3-s window centred on EqT picks to provide a priori estimate of arrival times for better accuracy (Zhang et al. 2003). Thus, we expect to obtain up to five picks for each target P and S phases. For quality control, we first use the Skewness and Z-test to remove outliers (far away from the rest), then we keep P and/or S phases with at least three picks with consistent arrivals, which are within a time window of 0.15 s for P phase and 0.2 s for S phase, respectively. Thus, based on the criteria, among those eligible P or S phases, they were all detected by EqT. At last, the optimal pick is determined by a function of Distance-Weighted Average (DWA) of the picks:

$$DWA = \frac{\sum_i^n W_i X_i}{\sum_i^n W_i} \quad (3)$$

where $X_i$ and $W_i$ correspond to $i^{th}$ pick and its corresponding weight:

$$W_i = \frac{1}{\sum |X_i - X_j|}, \quad i \neq j. \tag{4}$$

As an example, Fig. 3 shows how to determine optimal P and S phase labels based on a group of 4 picks.

To evaluate the performance of auto-labelling, we created a test dataset for benchmark, consisting of 2,983 P and 2,954 S arrivals obtained from 304 OBSs, which was a randomly selected subset of earthquake waveforms from all 11 networks (called sub-global test dataset). Using them as benchmark for performance evaluation, our auto-labelling approach outperforms the five popular phase pickers and generates the most reliable and accurate labels, with [Mean Absolute Error (MAE), standard deviation ($\sigma$), mean ($\mu$)] of [31, 60, -1] ms and [78, 99, 63] ms for P and S arrival times, respectively (Fig. 4). It demonstrates that our auto-labelling approach is effective for accurate and unbiased phase labelling. Among the five pickers, for P phases, AIC performed best with values of [39, 111, 10] ms, closely followed by PickNet with values of [45, 101, -25] s. For S phases, EqT demonstrated the best performance with values of [83, 102, -65] ms with a small margin to PickNet with values of [100, 163, 67] ms.

At last, we applied this automated labelling approach to the whole training dataset and labelled 35,829 60-s three-component earthquake seismograms, recorded by 423 OBSs with different deployment depths ranging from 0.05 km to 5.58 km (Figs. 1 and 5; Table S1). The labelled earthquake seismograms correspond to 1,974 events (out of 2,850 events) in the Alaska subduction zone, 231 events (out of 1,774 events) in the Cascadia subduction zone (above latitude 41.5 N), 1,014 events (out of 3,649 events) in the Gorda Plate (south of Cascadia, below latitude 41.5 N), 114 events (out of 276 events) in the Chile subduction zone, 3,888 events (out of 23,592 events) in the East-Pacific Rise, 190 events (out of 464 events) in the northern Mid-Atlantic Ridge region (southwest of Greenland), 24 events (out of 105 events) on the Hawaiian Ridge, 28 events (out of 135 events) on the San Andreas transform fault (offshore Los Angeles), 1,239 events (out of 4,958 events) in the New Zealand subduction zone, 347 events (out of 2,237 events) in the Taiwan subduction zone, and 129 events (out of 472 events) in the Venezuela subduction zone. The dataset is characterised by earthquake samples recorded at distances up to 352 km (Fig. 5a), with a wide range of reported depths from 0 to 234 km (with a majority of < 20 km; Fig. 5b), a wide range of magnitudes from 0 to 7 (various magnitude scales; Fig. 5c), and a broad distribution of SNRs (Fig. 5d).

## 2.3 Noise Auto-Extraction

In this study, real OBS noises are utilised in our model training and noisy data simulation. Here we utilise higher-order statistics (HOS), specifically Kurtosis, to extract seismic noise samples from continuous waveforms of the same 11 OBS networks. HOS enable us to classify signals as Gaussian or non-Gaussian distribution. As the HOS of a Gaussian distribution is equal to zero, the level of non-Gaussianity of a given signal is equivalent to its cumulants deviation from zero (Tsolis & Xenos 2011). Considering a zero mean seismic signal of N samples, the Kurtosis is defined as (Bickel & Doksum 1977; Mousavi & Langston 2016):

$$Kurtosis_y = \frac{\sum_{n=1}^{N}(y_n - \mu_y)^4}{N(\sigma_y^4)} - 3, \tag{5}$$

where $\mu_y$ and $\sigma_y$ are the mean and standard deviation of the signal y. The bias and variance for a Gaussian signal are given by $-\frac{6}{N}$ and $\frac{24}{N}$, respectively, and an estimator of this measure can be defined as:

$$|Kurtosis_y| \leq \frac{\sqrt{24/N}}{\sqrt{1-a}}, \tag{6}$$

in which *a* is an authorised confidence level with an optimal value of 90% (Ravier & Amblard 2001). Based on the Kurtosis function, we extract 25,000 noise samples from continuous waveforms recorded at the 11 OBS networks.

**2.4 Test Datasets**

To evaluate the performance of OBST, we conduct a comprehensive comparative analysis between OBST and EqT using three groups of distinct test datasets (Figs. 6). To assess the models' generalizability, we first adopted the sub-global test dataset (see Section 2.2) for performance evaluation. It is important to note that this dataset is additional to the dataset used for model training and testing. To simulate noisy earthquake waveforms under different ocean environments (e.g., different water depths, different tectonic settings, etc.), we superimpose two varying levels of high-frequency (i.e. > 3 Hz) OBS noise waveforms, which are randomly selected from the 25,000 OBS noise samples (see Section 2.3), onto the sub-global test dataset. Each three-component noisy earthquake waveform N corresponds to a combination of a three-component original earthquake waveform S and a three-component noise waveform ON:

$$N_i = S_i + (rand(a, b) * ON_i * \max(S_i)), \; i = 0,1,2 \tag{7}$$

where rand(a, b) is a random number between (0.05, 0.15) and (0.15, 0.25) for the two noisy datasets. These two specific noisy datasets are to evaluate the susceptibility of OBST against harsher noise levels. We refer to these sub-global test datasets as "original", "Noisy_L1", and "Noisy_L2" with increasing levels of noise.

The regional-scale test dataset consists of 5,047 earthquakes from the Alaska Amphibious Community Seismic Experiment (AACSE; Barcheck et al. 2020) starting from 12 May 2018 to 30 Jun 2019 (Ruppert et al. 2021a,b; see Fig. S1a). The ground truth labels comprise 20,829 P and 21,713 S samples with a maximum epicentral distance of 251 km and depths up to 215 km. The seismograms in this dataset are annotated following the below procedure (Ruppert et al. 2023): 1) running STA/LTA detector to determine the potential phases at all stations, 2) reviewing and fine-tuning the identified picks by student analysts, and 3) verified by supervising seismologists. The dataset and phase catalogue are publicly available from the IRIS Data Center under station code XO and UA@ScholarWorks publications (Ruppert et al. 2021a,b).

The local-scale test dataset consists of 3,649 earthquakes in the Gorda plate region (Guo et al. 2021; Fig. S1b). This dataset includes 18,095 P and 21,997 S arrivals, which were picked manually and fine-tuned by the AIC picker (Guo et al. 2021). The earthquake samples in this dataset have a depth range of 1 to 85 km and epicentral distances up to 112 km, providing insights into the model's performance at very local scales.

Here, a small fraction of earthquake samples from the AACSE and Gorda plate datasets are shared between the training process and performance evaluation procedure. ~13% earthquake samples of each AACSE and Gorda plate datasets were included in phase labelling based on the auto-labelling approach. Here we did not remove them from our following test datasets based on below three considerations: 1) we selected and relabelled all P and S picks for model training using the auto-labelling approach and did not adopt any of their manual picks, while performance evaluation is based on their manual picks; 2) those shared picks have to be picked by EqT based on the selection criteria in the auto-labelling process and are also most likely to be picked by the improved picker OBST in the performance evaluation procedure, playing a minor role in evaluating the recall performance. However, they may be important for precision evaluation because they mostly consist of high-quality phases in the dataset, with the most reliable manual picks; 3) we aim to evaluate the performance of OBST on typical earthquake catalogues as real cases in terms of distances and magnitudes, and the statistics distribution could be changed if we remove them (e.g. some local events are missing). Along with the three sub-global test datasets, we show the SNR distributions of earthquake samples of the AACSE and Gorda plate test datasets (Fig. 6). In section 5, the three groups of test datasets will be adopted for the performance comparison between the OBST and EqT under different noise levels, detection/picking thresholds, as well as distance and depth ranges.

## 4 Model Training

In this study, we leverage the transfer learning approach for model training. Transfer learning works based on the principle that different parts of the network learn distinct features. The earlier layers, typically convolutional layers, extract general components such as edges and simple textures, while the deeper ones learn more complex patterns related to the given task and dataset (Cunha et al. 2020). Several transfer learning strategies have been proposed, and a detailed survey can be found in Zhuang et al. (2020). Among them, the most common transfer learning strategy is "full fine-tuning", which has been shown to be an effective strategy in several seismological studies, outperforming training a model from scratch (e.g. Zini et al. 2020, Lapins et al. 2021). Under this strategy, the whole network parameters of a pre-trained model are modified, and "feature extractor", where the convolutional part of the pre-trained model is frozen (i.e., kept unchanged), and only the classification layers are retrained.

We adopt the "full fine-tuning" strategy to carry out the transfer learning using the EqT as base model. EqT has demonstrated superior performance compared to other DL phase pickers such as PhaseNet (Zhu and Beroza 2019), GPD (Ross et al. 2019), DeepPhasePick (DPP; Soto & Schurr 2021), and CNN-RNN Earthquake Detector (CRED; Mousavi, Zhu et al. 2019) in both event detection and phase picking tasks (Münchmeyer et al. 2021). The superior performance of EqT can be attributed to several factors: the quality and quantity of the training data, the attention mechanism, and the network's depth, which enhances the model's discriminatory power through learning more nonlinear mapping functions (Mousavi et al. 2020). Moreover, EqT was trained on the STEAD dataset using samples recorded up to 300 km away, making it potentially advantageous for OBS deployments with usually greater station spacing. Thus, we select EqT as our base model during our transfer learning. We randomly split our dataset into instances of training (70%), validation (15%), and test (15%) sets to ensure an even distribution of data in terms of classes.

Prior to training, we conducted an experiment to determine the impact of two factors on the stability and performance of our model: 1) the bandpass filtering used in data pre-processing and 2) the type of noise used for data augmentation. Usually raw waveform (e.g. PhaseNet) or filtered waveforms (e.g. 1-45 Hz in EqT) are adopted in model training for phase picking on land data. But we prefer to use a relatively narrower frequency range of 3-20 Hz for OBS data, which have been comparably used in other OBS studies (e.g. 3-20 Hz in Stone et al. 2018; 5-20 Hz in Ruppert et al. 2023), to eliminate the ocean-specific low-frequency sources of noise and lower the possibility of false detections in high frequencies. On the other hand, when using DL phase pickers filtered data also result in a better performance (e.g. Zhang et al. 2022; Chen et al. 2022). To compare their performance on the two frequency bands, we apply EqT on the "original" sub-global test dataset bandpass filtered between 1-45 Hz (the original frequency band in EqT) and 3-20 Hz (our preferred band). We calculate the evaluation metrics - recall (Rec), precision (Pre), and F1-score (F1) - that were used in Mousavi et al. (2020) to facilitate the performance comparison:

$$\Pr e = \frac{TP}{TP+FP}, Rec = \frac{TP}{TP+FN}, F1 = 2 \cdot \frac{\Pr e \cdot Rec}{\Pr e + Rec}. \tag{7}$$

We observed a boosted performance in phase arrival picking for the narrower frequency band, as evidenced by a 7% and 5% increase in F1-score values for P and S phases, respectively (Table 1). Whereas the performance drops significantly in terms of all metrics compared to the original paper of EqT, indicating that the waveform features are different between land and OBS recordings.

Using the optimal frequency band of 3-20 Hz and our training dataset, we trained two transfer-learned models following the later formal training procedure, differing only in the type of noise employed for data augmentation, i.e., Gaussian noise and real OBS noise. Tests show that adopting real OBS noise for model training results in higher probabilities for detection and P/S-wave picking in low SNR data (see examples in Fig. S2). Thus, we adopt the real OBS noise instead of the widely used Gaussian noise for data augmentation during the model training.

With the preferred frequency band of 3-20 Hz and real OBS noise, we adopt the EqT as our base model for transfer learning. The training was carried out several times to find the optimal hyperparameters, each time evaluating a single parameter. We train the model using Adam solver (Kingma & Ba 2014) to optimise the custom loss function used for the original EqT model (a weighted sum of 3 independent cross-entropy loss functions) and choose a learning rate of 1e-4 across the network. We consider a dropout rate of 0.2 and employ the early stop mechanism (Prechelt 2002) to avoid overfitting by monitoring the model performance over the validation set. The training was halted after 68 epochs as eight consecutive epochs without improvement in validation loss were met. The data was augmented during model training by adding OBS noise and gaps, dropping channels, and shifting the events through the signals (Mousavi et al. 2020). We set the P and S phase labels as Gaussian windows of 40-sample (0.4 s) length. Additionally, we define the length of coda waves as 1.2 times P and S travel time difference starting from the S phase, which helps discard the lower-energy content of the S coda that is more challenging for the model to learn.

# 5 Performance Evaluation

To conduct the performance comparison between the two pickers, we follow the convention of Mousavi et al. (2020) and define the picking residuals as the difference between ground truth and model picks. Then, we analyse the metrics analysis (F1-score, precision, and recall) and statistical analysis ($\mu$, $\sigma$, and MAE) to quantify and compare the pickers' performance. In the metrics analysis, we consider an error tolerance of 0.5 s (i.e. effective detection window) for P and S phases across all tests except for the S-phase in the Section 5.2. In the statistics analysis, we exclude outliers from the calculation to improve the clarity of the residual distributions (Bornstein et al. 2023) and the outlier window is determined based on the frequency distribution of the P and S residuals in each test.

The input data in the following tests are one-minute three-component seismograms associated with one catalogued earthquake, although multiple undocumented events may be present in the samples. To account for this possibility, we evaluate all P and S detections within each input sample and consider the closest detection to the corresponding benchmark as either a TP or FP. If the P or S benchmark does not fall within the detection windows, it is classified as an FN.

## 5.1 Comparison on Sub-Global Test Dataset

As for generalizability assessment, we compare the performance of the two models using our manually labelled sub-global test datasets, which are from the same 11 OBS deployments in the training dataset. The datasets consist of the original sub-global test dataset (i.e., "original") and its two noisy variants (Noisy_L1 and Noisy_L2). The two noisy test datasets are designed to investigate the pickers' abilities in harsher noise levels. We acknowledge that a comparison on such small datasets may not fully reflect the generalizability of the models, but this evaluation can still offer valuable insights into their performance.

We comprehensively compare the performance of the OBST and EqT across the three test datasets with different noise levels. We used two different detection/picking thresholds: 0.5/0.3, which was originally adopted by Mousavi et al. 2020, and 0.3/0.1, which was widely used in practical studies (e.g. Jiang et al. 2021; Chen et al. 2022). Here we set the effective detection window as 0.5 s in the metrics analysis and consider the picks with residuals larger than 1 s as outliers in the statistical analysis (Fig. 7; Table S2). The matrices of F1-score, precision, and recall rates are summarized in Tables 2-4. The results show that OBST consistently outperforms EqT across all noise levels, especially at higher thresholds. OBST improves precision and recall rates for P onsets, up to 4% and 15%, respectively, with a narrower range of P residuals (Fig. 7). While for S arrivals both pickers exhibit comparable precisions (Tables 2-4) and residual distributions (identified by $\sigma$ values in Fig. 7). But the recall rate of S phases consistently improves at different noise levels and thresholds, up to 32% in the Noisy_L2 dataset at higher thresholds. We compare the average MAEs for P arrivals at [lower, higher] thresholds across the three test datasets, our OBST results in half MAE ([71 ms, 63ms]) compared to the EqT [136 ms, 128 ms]). To fairly compare the precision of the two pickers, we count those common events that can be picked by both pickers (see the insets in each panel in Fig. 7). OBST consistently outperforms EqT regarding P-wave precision across all scenarios, while S precision is the same or comparable.

We also investigate the impact of noise level and prediction/picking thresholds on model performance in terms of TP, FP, and FN rates (Fig. 8). Across the three noise levels, for OBST, TP differences between the two thresholds are [30, 64, 196] for P phases (Fig. 8a) and [27, 72, 209] for S phases (Fig. 8d). In contrast, for EqT, the same criterion produces higher values: [188, 218, 344] for P phases and [435, 443, 534] for S phases. Thus, OBST is less reliant on the choice of detection/picking thresholds. In addition, OBST exhibits less FPs for P phases across all investigated scenarios (Fig. 8b). Notably, in higher thresholds, EqT displays a decreasing trend in FP for P-phase as the noise level increases, which coincides with a noticeable rise in its FNs. These observations suggest that the reduction in EqT's FPs is primarily due to the loss of noisy and/or complex seismograms, leaving only high-quality samples available. On the other hand, OBST experiences a modest decrease in P precision (slight increase in FPs) as a trade-off for maintaining a low rate of FNs. We speculate that the overall stability of OBST against varying noise levels is due primarily to its capacity to produce higher probabilities for both P and S detections in noisy conditions (see Fig. S3). This is coupled with the model's ability to keep residuals reliably low. These findings suggest that OBST has learned to extract subtle features from the OBS data that EqT could not capture, which may be partly attributed to employing OBS noise for data augmentation instead of Gaussian noise in OBST model training.

## 5.2 Comparison on Regional-Scale Dataset

Ruppert et al. (2023) investigated EqT's performance on continuous waveforms from 55 OBS stations collected over 14 months spanning from June 1, 2018, to July 31, 2019. They pointed out that EqT was most effective at depths ranging from 40 to 80 km and distances less than 100 km. However, the overall performance on OBS data was considerably inferior to that of STEAD. In this section, we evaluate and compare the performance of OBST with EqT using one-minute seismograms of 5,047 events from the AACSE dataset across three different distance and depth ranges. We choose an effective detection window of 0.5 seconds for the P phase as before but 1 second for the S phase. This decision is motivated by the large uncertainty standard deviation associated with S picks in the AACSE catalogue (Fig. S4; Table S3). In our statistical analysis, we consider picks with residuals larger than 2 seconds as outliers. This determination is made based on the frequency distribution of S residuals observed in Figures 9 and 10. For this test and the one in Section 5.3, we experimentally set the detection and P/S picking thresholds to 0.3 and 0.1.

We first evaluate the performance of OBST and EqT at various epicentral distances (Fig. 9; Table 5). OBST demonstrates a stable and consistent capability for P and S phase picking at different epicentral distances, as indicated by F1-score values (Table 5). In contrast, EqT's performance significantly drops, with F1-scores plummeting from 0.82 to 0.14 for P-phase and from 0.88 to 0.12 for S-phase, due mostly to an abundance of missing picks. The OBST shows the maximum improvement of recall rates for P and S phases at farther distances: 68% for P picks and 76% for S picks. The P and S recall rates of EqT are significantly decreased with the increasing of distances, most likely due to the longer complex source-receiver wave paths passing through subduction zone structures. At different distances, the precisions of P and S picks for OBST and EqT are comparable, within difference of several percentages (Table 5). At distances greater than 200 km, EqT outperforms OBST by 6% in S precision. However, EqT only detects 8% of OBST's detections (Table 5). A fair way to compare the precision is to analysis common picks by the two pickers. Using those common picks as references, OBST performs slightly better precision for P picks and still comparable for S picks (inset in Fig. 9). Overall, our OBST does a much better job for large distances, especially for distances of > 200 km, demonstrating its success to handle low SNR data (Fig. 9k,l).

We evaluate the performance of OBST and EqT at various depth ranges (Fig. 10; Table 6). OBST demonstrates consistent high P and S recall rates across all depth ranges, with values of 77%/86% (<20 km), 82%/89% (20 - 80 km), and 77%/83% (>80 km) while keeping the precision comparable to the EqT (Table 6). Conversely, the recall rate of EqT is very sensitive to depth. EqT results in P and S recall rates of 47% and 48% at depths less than 20 km and 57% and 56% at depths of 20 - 80 km, dropping to 35% and 36% at depths greater than 80 km (Table 6). Similar to the distance-based comparison, comparison on common detections demonstrate that OBST leads to higher P-pick precision (insets in Fig. 10a,e,i) and comparable S-pick precision (insets in Fig. 10b,f,i). The average SNR of the detected P and S phases declines as the waves travel through great depths (e.g. >80 km; Fig. 10k,l).

For the P phases, the average MAEs of OBST/EqT are 142/137 ms over 16,709/10,858 detections with an average SNR of 9.2/10.7 db. For the average MAEs of S phases, EqT performs slightly better than OBST with a 26 ms discrepancy over 11,380 detections (EqT) vs 19,148 detections (OBST). While we adopt a reserved stance when discussing the precision, especially for S phases. We manually verified some P and S FPs by both pickers. We noticed that some P and S picks were labelled inaccurately in the AACSE catalogue (see examples in Figs. S5-6). Furthermore, we observed inconsistent arrivals of S phases on two horizontal components (refer to the discussions in Section 6.2). However, only one S arrival can be labelled and predicted with randomness, which may affect the performance evaluation.

### 5.3 Comparison on Local-Scale Dataset

To evaluate the performance of OBST at very local distances, we applied the model on one-minute waveforms of catalogued earthquakes in the Gorda plate region and compared the results with those of EqT. The highly complex tectonic settings in this region, particularly along the Mendocino Triple Junction, presents an ideal testbed for this comparison. In the matrices analysis, we choose the effective detection window as 0.5 seconds for both P and S phases. In the statistical analysis, we consider picks with residuals larger than 1 second as outliers, which is based on the frequency distribution of P residuals (Figs. 11 and 12). Overall, OBST outperforms EqT in terms of F1-score for both P and S phases across all distances, as shown in Table 7. Specifically, OBST performs best in P-wave precision at different distances, outperforming EqT by up to 7% (0.99 - OBST vs. 0.92 - EqT), which are applicable to those common picks as well (Figs. 11a,e,i). In contrast, both OBST and EqT perform consistently well in S precision (up to 99%), within a slight difference of 1%, for both the whole dataset and common picks (Figs. 11b,f,j). Systematic precedence of approximately 100 ms can be observed in S picks by both pickers, which are also observed for those common detections (the insets in Figs. 12b,f,j). Upon investigating the corresponding seismograms, we speculate that this systematic shift is likely due to the fine-tuning step performed by the AIC picker following manual arrival labelling (Guo et al. 2021), which is particularly challenging for S phases.

Regarding the comparison at different depth ranges (Fig. 12), the residual distributions are similar to that at different distance ranges. OBST shows narrower distributions of P-phase residuals than EqT consistently at all depth ranges. Both pickers demonstrate relatively good performance at various depths (Fig. 12). For P phases, the best performance is at the depth range of 10-20 km for OBST and at the depths greater than 20 km for EqT, with the [MAE, $\sigma$, $\mu$] values of [63, 78, 50] and [191, 246, 73] ms respectively. For S phases, the best performance is at the depths greater than 20 km with the [MAE, $\sigma$, $\mu$] values of [80, 78, 71] ms for OBST and [67, 78, 51] ms for EqT. OBST outperforms EqT in terms of recall rates for P and S picks (Table 8). This improvement is maximum at depths greater than 20 km by 15% (P) and 20% (S). Moreover, OBST excels in P precisions by 7% (depth < 10 km), 7% (10 km < depth < 20 km), and 3% (depth > 20 km) with increasing depth (Fig. 12). The improvement of both recall rates and precisions indicates the additional phase detections by OBST are of high precision. As an example, we compare the P and S phase picking using OBST and EqT along with the sub-global and regional-scale tests (Fig. 13). OBST generates higher detection and picking probabilities than EqT, leading to more reliable and precise phase picks.

## 6 Discussions

### 6.1 Auto-Labelling and Model Performance

In this study, we proposed a novel approach for labelling training datasets, utilising DL and conventional phase pickers. In this method, a label is assigned based on the agreement of at least N collected picks within a designated time window $\Delta t$. A shorter time window may result in more accurate labels yet lead to fewer labelled samples for training due to fewer consistent picks. Conversely, a longer time window may provide more consistent picks and more labelled samples, but may also result in less precise labels. Therefore, balancing these parameters based on dataset quality and the desired precision is critical to obtain the most accurate and substantial labelled dataset. Based on our experiments, we found the optimal parameters as an agreement of at least three picks within a time window of 0.15 and 0.2 s for P and S phases, respectively. This set of parameters delivered an excellent performance on our sub-global test dataset (i.e., "original" test dataset), generating labels for 58% of the samples with MAE, $\sigma$, and $\mu$ of 31, 60, and -1 ms for P phases and 78, 99, and 63 ms for S phases.

To visually inspect the impact of transfer learning on EqT using our training dataset, we compare the evolution of feature maps (i.e., the output of a set of filters at a given CNN layer) inside the P-decoder of OBST and EqT (Fig. 14). The fluctuations in the feature maps with different amplitudes suggest that the models can detect various features at different levels of abstraction. As the decoder layers progress (from top to bottom, Fig. 14), the flattening of some feature maps implies that those particular features were less relevant or important to the designated task. In contrast, the feature maps that point to the actual P arrival time as a sharp pick or concavity are likely the most relevant for accurate P-phase picking. The feature maps in OBST accurately point at the P arrival time, resulting in P-detection with a probability of 94% and a residual of 80 ms, as evidenced in the last three CNN layers. On the other hand, in EqT, the feature maps tend to introduce a relatively wider range of samples with different probabilities as candidates for the P arrival time. If a picking probability threshold of > 0.14 or an effective detection window of < 2.12 s are used, this can result in false negative or false positive.

## 6.2 S-wave Picking Challenges on OBS Data

Labelling P and S arrivals in OBS data can be challenging, requiring expertise and experience in seismic data analysis. The difficulty level can depend on various factors, such as the quality of the data (e.g. noise level) and the complexity of the seismic waveforms (e.g. wave scattering). The P-wave arrival is relatively easier to identify, as it is the first arrival on the seismogram and has a distinctive onset. However, the identification and picking of S-wave can be more challenging, as it can be masked by other seismic phases, such as multiples or converted phases. What's more, S arrivals recorded on two horizontal components can be inconsistent under certain circumstances. Our observations reveal the existence of inconsistent S-wave arrivals on the two horizontal components, making picking judgement challenging (Fig. S7). One possible explanation is the presence of anisotropic sediments, which leads to shear wave splitting (Volti et al. 2005), posing a challenge for phase picking using DL models.

## 6.4 Recommendations

OBST is an efficient DL model that greatly enhances phase picking performance on OBS data. Among the existing DL phase pickers, it stands out as a user-friendly tool with straightforward parameter settings. Similar to the recently released PickBlue (the pre-print version of the paper is available at Bornstein et al. 2023), OBST can be employed independently or seamlessly integrated into established seismology toolboxes such as Seisbench (Woolam et al. 2022). However, constructing an accurate earthquake catalogue still relies on conventional earthquake association and location procedures. When dealing with extensive OBS data, we recommend combining OBST with established seamless earthquake location workflows, such as LOC-FLOW (Zhang et al. 2022). This integration would provide the users with the utmost benefits in terms of accessibility and convenience.

The performance of OBST exhibits variations across different distance ranges and depth ranges, which is influenced not only by the availability of corresponding labels but also by the diverse tectonic settings. Subduction zones, for instance, experience lateral and vertical seismic complexities due to the intricate processes associated with subduction. To enhance the performance of OBST in specific local studies, an additional round of transfer learning can be beneficial if well-labelled samples are accessible. This process allows the model to leverage the knowledge gained from pre-existing training on general data and apply it to the specific local context, thereby improving its accuracy and effectiveness.

## 7 Conclusion

We utilized seismic data collected from 11 temporary OBS deployments worldwide to create a relatively small but suitable dataset for transfer learning, comprising 35,829 earthquakes and 25,000 noise samples with a broad range of distances and depths. By employing popular deep learning phase pickers (EqT, PhaseNet, GPD, PickNet) and the AIC picker, we introduced a novel approach for automatically labelling training datasets. Then we train our OBS phase picker - OBST using the well-trained EqT model as the base model and the automatically labelled phases as the training dataset through transfer learning. Notably, for data augmentation, we opted to utilize real OBS noise samples instead of the more commonly employed Gaussian noise. By comparing with the EqT, our OBST demonstrates a significant improvement in phase detection and arrival picking, particularly in terms of recall rates, as evidenced by extensive testing and comparisons. In extreme cases from the AACSE dataset with epicentral distances exceeding 200 km, OBST outperformed EqT with increased recall rates of 68% and 76% for P- and S-phases, respectively. Whereas improvements in P and S precision were relatively small or similar, which may be influenced by the accuracy of training and benchmarked labels, especially for S phases.

**Acknowledgements**

The authors sincerely appreciate developers and providers of deep-learning models and seismic phase catalogues used in this study, which were cited accordingly in the main text. In addition, we thank the associate editor xx and the reviewers xx for their valuable comments. This work is supported by the Natural Sciences and Engineering Research Council of Canada Discovery Grant (RGPIN-2019-04297). This research was enabled in part by support provided by Canada Foundation for Innovation.

**Conflict of interest statement**

The authors declare no conflicts of interest related to this work.

**Data Availability Statement**

The earthquake catalogues and waveforms were downloaded from the Data Management Center at the Incorporated Research Institutions for Seismology (https://ds.iris.edu/ds/nodes/dmc/, last accessed in May 2023) using the ObsPy package (Beyreuther et al. 2010; Krischer et al. 2015). The earthquake catalogues specifically for the Gorda plate and East Pacific Rise were sourced from Guo et al. (2021) and Gong et al. (2022) respectively. For performance evaluation at local and regional scales, manual picks from Guo et al. (2021) for the Gorda plate and from Ruppert et al. (2021a,b) for the Alaska Amphibious Community Seismic Experiment were utilized. OBSTransformer is publicly available and maintained at https://github.com/alirezaniki/OBSTransformer. It is worth noting that the model will soon be integrated into LOC-FLOW as well (Zhang et al. 2022).

**References**

Allen, R. V, 1978. Automatic earthquake recognition and timing from single traces. *Bulletin of the Seismological Society of America*, **68**, 1521-1532.

Barcheck, G., Abers, G.A., Adams, A.N., Bécel, A., Collins, J., Gaherty, J.B., Haeussler, P.J., *et al.*, 2020. The Alaska amphibious community seismic experiment. *Seismological Research Letters*, **91**, 3054–3063, Seismological Society of America. doi:10.1785/0220200189

Beyreuther, M., Barsch, R., Krischer, L., Megies, T., Behr, Y. & Wassermann, J., 2010. ObsPy: A python toolbox for seismology. *Seismological Research Letters*, **81**, 530–533. doi:10.1785/gssrl.81.3.530

Bickel, P.J. & Doksum, K.A., 1977. Mathematical Statistics Holden-Day. *Inc., SF*.

Bornstein, T., Lange, D., Münchmeyer, J., Woollam, J., Rietbrock, A., Barcheck, G., Grevemeyer, I., *et al.*, 2023. PickBlue: Seismic phase picking for ocean bottom seismometers with deep learning. Retrieved from http://arxiv.org/abs/2304.06635

Chai, C., Maceira, M., Santos-Villalobos, H.J., Venkatakrishnan, S. V., Schoenball, M., Zhu, W., Beroza, G.C., *et al.*, 2020. Using a Deep Neural Network and Transfer Learning to Bridge Scales for Seismic Phase Picking. *Geophys Res Lett*, **47**, Blackwell Publishing Ltd. doi:10.1029/2020GL088651

Chen, H., Yang, H., Zhu, G., Xu, M., Lin, J. & You, Q., 2022. Deep Outer-Rise Faults in the Southern Mariana Subduction Zone Indicated by a Machine-Learning-Based High-Resolution Earthquake Catalog. *Geophys Res Lett*, **49**, John Wiley and Sons Inc. doi:10.1029/2022GL097779

Cunha, A., Pochet, A., Lopes, H. & Gattass, M., 2020. Seismic fault detection in real data using transfer learning from a convolutional neural network pre-trained with synthetic seismic data. *Comput Geosci*, **135**, Elsevier Ltd. doi:10.1016/j.cageo.2019.104344

Gong, J., Fan, W. & Parnell-Turner, R., 2022, February 16. Microseismicity Indicates Atypical Small-Scale Plate Rotation at the Quebrada Transform Fault System, East Pacific Rise. *Geophys Res Lett*, John Wiley and Sons Inc. doi:10.1029/2021GL097000

Guo, H., McGuire, J.J. & Zhang, H., 2021. Correlation of porosity variations and rheological transitions on the southern Cascadia megathrust. *Nat Geosci*, **14**, 341–348, Nature Research. doi:10.1038/s41561-021-00740-1

Hochreiter, S. & Schmidhuber, J., 1997. Long short-term memory. *Neural Comput*, **9**, 1735–1780, MIT press.

Jiang, C., Fang, L., Fan, L. & Li, B., 2021. Comparison of the earthquake detection abilities of PhaseNet and EQTransformer with the Yangbi and Maduo earthquakes. *Earthquake Science*, **34**, 425–435, Earthquake Science. doi:10.29382/eqs-2021-0038

Kingma, D.P. & Ba, J., 2014. Adam: A Method for Stochastic Optimization. Retrieved from http://arxiv.org/abs/1412.6980

Krischer, L., Megies, T., Barsch, R., Beyreuther, M., Lecocq, T., Caudron, C. & Wassermann, J., 2015. ObsPy: A bridge for seismology into the scientific Python ecosystem. *Comput Sci Discov*, **8**, Institute of Physics Publishing. doi:10.1088/1749-4699/8/1/014003

Lapins, S., Goitom, B., Kendall, J.M., Werner, M.J., Cashman, K. V. & Hammond, J.O.S., 2021. A Little Data Goes a Long Way: Automating Seismic Phase Arrival Picking at Nabro Volcano With Transfer Learning. *J Geophys Res Solid Earth*, **126**, John Wiley and Sons Inc. doi:10.1029/2021JB021910

LeCun, Y., Bengio, Y. & others, 1995. Convolutional networks for images, speech, and time series. *The handbook of brain theory and neural networks*, **3361**, 1995, Citeseer.

Liu, C., Ke, W., Jiao, J. & Ye, Q., 2017. RSRN: Rich side-output residual network for medial axis detection. Proceedings–2017 IEEE International Conference on Computer Vision Workshops, ICCVW 2017, 1739–1743.

Luong, M.-T., Pham, H. & Manning, C.D., 2015. Effective Approaches to Attention-based Neural Machine Translation. Retrieved from http://arxiv.org/abs/1508.04025

Maeda, N., 1985. A method for reading and checking phase times in autoprocessing system of seismic wave data. *Zisin*, **38**, 365–379.

Mousavi, S.M. & Langston, C.A., 2016. Hybrid seismic denoising using higher-order statistics and improved wavelet block thresholding. *Bulletin of the Seismological Society of America*, **106**, 1380–1393, Seismological Society of America. doi:10.1785/0120150345

Mousavi, S.M., Ellsworth, W.L., Zhu, W., Chuang, L.Y. & Beroza, G.C., 2020. Earthquake transformer—an attentive deep-learning model for simultaneous earthquake detection and phase picking. *Nat Commun*, **11**, Nature Research. doi:10.1038/s41467-020-17591-w

Mousavi, S.M., Sheng, Y., Zhu, W. & Beroza, G.C., 2019. STanford EArthquake Dataset (STEAD): A Global Data Set of Seismic Signals for AI. *IEEE Access*, **7**, 179464–179476, Institute of Electrical and Electronics Engineers Inc. doi:10.1109/ACCESS.2019.2947848

Mousavi, S.M., Zhu, W., Sheng, Y. & Beroza, G.C., 2019. CRED: A Deep Residual Network of Convolutional and Recurrent Units for Earthquake Signal Detection. *Sci Rep*, **9**, Nature Publishing Group. doi:10.1038/s41598-019-45748-1

Münchmeyer, J., Woollam, J., Rietbrock, A., Tilmann, F., Lange, D., Bornstein, T., Diehl, T., *et al.*, 2021. Which picker fits my data? A quantitative evaluation of deep learning based seismic pickers. doi:10.1029/2021JB023499

Ni, Y., Hutko, A., Skene, F., Denolle, M., Malone, S., Bodin, P., Hartog, R., *et al.*, 2023. Curated Pacific Northwest AI-ready Seismic Dataset. *Seismica*, **2**. doi:10.26443/seismica.v2i1.368

Obana, K., Scherwath, M., Yamamoto, Y., Kodaira, S., Wang, K., Spence, G., Riedel, M., *et al.*, 2015. Earthquake activity in northern Cascadia subduction zone off vancouver island revealed by ocean-bottom seismograph observations. *Bulletin of the Seismological Society of America*, **105**, 489–495, Seismological Society of America. doi:10.1785/0120140095

Prechelt, L., 2002. Early stopping-but when? in *Neural Networks: Tricks of the trade*, pp. 55–69, Springer.

Ravier, P. & Amblard, P.O., 2001. Wavelet packets and de-noising based on higher-order-statistics for transient detection. *Signal Processing*, **81**, 1909–1926. doi:10.1016/S0165-1684(01)00088-3

Ross, Z.E., Meier, M.A., Hauksson, E. & Heaton, T.H., 2018. Generalized seismic phase detection with deep learning. *Bulletin of the Seismological Society of America*, **108**, 2894–2901, Seismological Society of America. doi:10.1785/0120180080

Ruppert, N., Barcheck, G. & Abers, G., 2021. AACSE earthquake catalog: January-August, 2019.

Ruppert, N.A., Barcheck, G. & Abers, G.A., 2021. AACSE earthquake catalog: May-December, 2018.

Ruppert, N.A., Barcheck, G. & Abers, G.A., 2023. Enhanced Regional Earthquake Catalog with Alaska Amphibious Community Seismic Experiment Data. *Seismological Research Letters*, **94**, 522–530, Seismological Society of America. doi:10.1785/0220220226

Saragiotis, C.D., Hadjileontiadis, L.J. & Panas, S.M., 2002. PAI-S/K: A robust automatic seismic P phase arrival identification scheme. *IEEE Transactions on Geoscience and Remote Sensing*, **40**, 1395–1404. doi:10.1109/TGRS.2002.800438

Scherwath, M., Spence, G., Obana, K., Kodaira, S., Wang, K., Riedel, M., McGuire, J., *et al.*, 2011. Seafloor seismometers monitor northern Cascadia earthquakes. *Eos, Transactions American Geophysical Union*, **92**, 421–422, Wiley Online Library.

Simonyan, K. & Zisserman, A., 2014. Very Deep Convolutional Networks for Large-Scale Image Recognition. Retrieved from http://arxiv.org/abs/1409.1556

Sleeman, R. & Eck, T. Van, 1999. Robust automatic P-phase picking: an on-line implementation in the analysis of broadband seismogram recordings. *Physics of the Earth and Planetary Interiors*, Vol. 113.

Soto, H. & Schurr, B., 2021. DeepPhasePick: A method for detecting and picking seismic phases from local earthquakes based on highly optimized convolutional and recurrent deep neural networks. *Geophys J Int*, **227**, 1268–1294, Oxford University Press. doi:10.1093/gji/ggab266

Stone, I., Vidale, J.E., Han, S. & Roland, E., 2018. Catalog of Offshore Seismicity in Cascadia: Insights Into the Regional Distribution of Microseismicity and its Relation to Subduction Processes. *J Geophys Res Solid Earth*, **123**, 641–652, Blackwell Publishing Ltd. doi:10.1002/2017JB014966

Tan, Y.J., Waldhauser, F., Ellsworth, W.L., Zhang, M., Zhu, W., Michele, M., Chiaraluce, L., *et al.*, 2021. Machine-Learning-Based High-Resolution Earthquake Catalog Reveals How Complex Fault Structures Were Activated during the 2016–2017 Central Italy Sequence. *The Seismic Record*, **1**, 11–19, Seismological Society of America (SSA). doi:10.1785/0320210001

Tsolis, G. & Xenos, T.D., 2011. Signal Denoising Using Empirical Mode Decomposition and Higher Order Statistics Beamforming and Direction of Arrival Estimation Techniques View project Mobility and training for beyond 5G ecosystems [MOTOR5G] (HORIZON 2020 Marie Skłodowska-Curie Innovative Training Networks) View project SEE PROFILE Signal Denoising Using Empirical Mode Decomposition and Higher Order Statistics. *International Journal of Signal Processing*, Vol. 4. Retrieved from

https://www.researchgate.net/publication/228522183

Vaswani, A., Shazeer, N., Parmar, N., Uszkoreit, J., Jones, L., Gomez, A.N., Kaiser, L., *et al.*, 2017. Attention Is All You Need. Retrieved from http://arxiv.org/abs/1706.03762

Volti, T., Kaneda, Y., Zatsepin, S. & Crampin, S., 2005. An anomalous spatial pattern of shear-wave splitting observed in Ocean Bottom seismic data above a subducting seamount in the Nankai Trough. *Geophys J Int*, **163**, 252–264. doi:10.1111/j.1365-246X.2005.02743.x

Wang, J., Xiao, Z., Liu, C., Zhao, D. & Yao, Z., 2019. Deep Learning for Picking Seismic Arrival Times. *J Geophys Res Solid Earth*, **124**, 6612–6624, Blackwell Publishing Ltd. doi:10.1029/2019JB017536

Webb, S.C., 1998. Broadband seismology and noise under the ocean. *Reviews of Geophysics*, **36**, 105–142, Blackwell Publishing Ltd. doi:10.1029/97RG02287

Woollam, J., Münchmeyer, J., Tilmann, F., Rietbrock, A., Lange, D., Bornstein, T., Diehl, T., *et al.*, 2022. SeisBench-A Toolbox for Machine Learning in Seismology. *Seismological Research Letters*, **93**, 1695–1709, Seismological Society of America. doi:10.1785/0220210324

Xiao, Z., Wang, J., Liu, C., Li, J., Zhao, L. & Yao, Z., 2021. Siamese Earthquake Transformer: A Pair-Input Deep-Learning Model for Earthquake Detection and Phase Picking on a Seismic Array. *J Geophys Res Solid Earth*, **126**, Blackwell Publishing Ltd. doi:10.1029/2020JB021444

Zhang, H., Thurber, C. & Rowe, C., 2003. Automatic P-wave arrival detection and picking with multiscale wavelet analysis for single-component recordings. *Bulletin of the Seismological Society of America*, **93**, 1904–1912, Seismological Society of America. doi:10.1785/0120020241

Zhang, M., Liu, M., Feng, T., Wang, R. & Zhu, W., 2022. LOC-FLOW: An End-to-End Machine Learning-Based High-Precision Earthquake Location Workflow. *Seismological Research Letters*, **93**, 2426–2438, Seismological Society of America. doi:10.1785/0220220019

Zhao, M., Xiao, Z., Chen, S. & Fang, L., 2023. DiTing: A large-scale Chinese seismic benchmark dataset for artificial intelligence in seismology. *Earthquake Science*, **36**, 84–94. doi:10.1016/j.eqs.2022.01.022

Zhou, Y., Yue, H., Zhou, S. & Kong, Q., 2019. Hybrid event detection and phase-picking algorithm using convolutional and recurrent neural networks. *Seismological Research Letters*, **90**, 1079–1087, Seismological Society of America. doi:10.1785/0220180319

Zhu, J., Li, Z. & Fang, L., 2022. USTC-Pickers: a Unified Set of seismic phase pickers Transfer learned for China. *Earthquake Science*, **36**, 1–11, Earthquake Science.

Zhu, L., Peng, Z., McClellan, J., Li, C., Yao, D., Li, Z. & Fang, L., 2019. Deep learning for seismic phase detection and picking in the aftershock zone of 2008 Mw7.9 Wenchuan Earthquake. *Physics of the Earth and Planetary Interiors*, **293**, Elsevier B.V. doi:10.1016/j.pepi.2019.05.004

Zhu, W. & Beroza, G.C., 2019. PhaseNet: A deep-neural-network-based seismic arrival-time picking method. *Geophys J Int*, **216**, 261–273, Oxford University Press. doi:10.1093/gji/ggy423

Zhu, W. & Beroza, G.C., 2019. PhaseNet: A deep-neural-network-based seismic arrival-time picking method. *Geophys J Int*, **216**, 261–273, Oxford University Press. doi:10.1093/gji/ggy423

Zhuang, F., Qi, Z., Duan, K., Xi, D., Zhu, Y., Zhu, H., Xiong, H., *et al.*, 2021, January 1. A Comprehensive Survey on Transfer Learning. *Proceedings of the IEEE*, Institute of Electrical and Electronics Engineers Inc. doi:10.1109/JPROC.2020.3004555

Zini, J. El, Rizk, Y. & Awad, M., 2020. A Deep Transfer Learning Framework for Seismic Data Analysis: A Case Study on Bright Spot Detection. *IEEE Transactions on Geoscience and Remote Sensing*, **58**, 3202–3212, Institute of Electrical and Electronics Engineers Inc. doi:10.1109/TGRS.2019.2950888

Table 1. Performance of EqT on 1-45 Hz and 3-20 Hz filtered OBS data from the "Original" sub-global test dataset.

| Band (Hz) | P_Pre | P_Rec | P_F1 | S_Pre | S_Rec | S_F1 |
|---|---|---|---|---|---|---|
| 1-45 | 0.95 | 0.76 | 0.84 | 0.99 | 0.71 | 0.83 |
| 3-20 | 0.95 | 0.87 | 0.91 | 1 | 0.79 | 0.88 |

Abbreviations: Pre = Precision, Rec = Recall, F1 = F1-score

Table 2. Performance comparison on the "Original" sub-global test dataset at two different detection/picking thresholds.

| Model | P_Pre | P_Rec | P_F1 | S_Pre | S_Rec | S_F1 |
|---|---|---|---|---|---|---|
| Detection and picking thresholds are 0.5 and 0.3, respectively | | | | | | |
| OBST | 0.99 | 0.95 | 0.96 | 0.99 | 0.95 | 0.97 |
| EqT | 0.95 | 0.87 | 0.91 | 1 | 0.79 | 0.88 |
| Detection and picking thresholds are 0.3 and 0.1, respectively | | | | | | |
| OBST | 0.98 | 0.96 | 0.97 | 0.99 | 0.97 | 0.98 |
| EqT | 0.95 | 0.94 | 0.94 | 0.99 | 0.95 | 0.97 |

Abbreviations: OBST = OBSTransformer, Pre = Precision, Rec = Recall, F1 = F1-score

Table 3. Performance comparison on the "Noisy_L1" sub-global test dataset at two different detection/picking thresholds.

| Model | P_Pre | P_Rec | P_F1 | S_Pre | S_Rec | S_F1 |
|---|---|---|---|---|---|---|
| Detection and picking thresholds are 0.5 and 0.3, respectively | | | | | | |
| OBST | 0.99 | 0.91 | 0.95 | 0.99 | 0.92 | 0.95 |
| EqT | 0.96 | 0.83 | 0.89 | 1 | 0.76 | 0.86 |
| Detection and picking thresholds are 0.3 and 0.1, respectively | | | | | | |
| OBST | 0.98 | 0.94 | 0.96 | 0.99 | 0.95 | 0.97 |
| EqT | 0.95 | 0.92 | 0.93 | 0.99 | 0.92 | 0.95 |

Abbreviations: OBST = OBSTransformer, Pre = Precision, Rec = Recall, F1 = F1-score

Table 4. Performance comparison on the "Noisy_L2" sub-global test dataset at two different detection/picking thresholds.

| Model | P_Pre | P_Rec | P_F1 | S_Pre | S_Rec | S_F1 |
|---|---|---|---|---|---|---|
| Detection and picking thresholds are 0.5 and 0.3, respectively | | | | | | |
| OBST | 0.97 | 0.66 | 0.78 | 0.97 | 0.82 | 0.89 |
| EqT | 0.95 | 0.51 | 0.66 | 0.99 | 0.50 | 0.66 |
| Detection and picking thresholds are 0.3 and 0.1, respectively | | | | | | |
| OBST | 0.93 | 0.75 | 0.83 | 0.97 | 0.89 | 0.93 |
| EqT | 0.91 | 0.65 | 0.76 | 0.98 | 0.69 | 0.81 |

Abbreviations: OBST = OBSTransformer, Pre = Precision, Rec = Recall, F1 = F1-score

**Table 5.** Performance comparison on the AACSE regional-scale test dataset at three different epicentral distance ranges.

| | Performance comparison in epicentral distances less than 100 km | | | | | |
|---|---|---|---|---|---|---|
| **Model** | **P_Pre (TP/FP)** | **P_Rec (FN)** | **P_F1** | **S_Pre (TP/FP)** | **S_Rec (FN)** | **S_F1** |
| **OBST** | 0.94 (4,888/300) | 0.80 (1,190) | 0.87 | 0.93 (4,875/341) | 0.91 (495) | 0.92 |
| **EqT** | 0.92 (4,440/386) | 0.74 (1,552) | 0.82 | 0.94 (4,451/297) | 0.82 (963) | 0.88 |
| | Performance comparison in epicentral distances greater than 100 km but less than 200 km | | | | | |
| **OBST** | 0.93 (7,538/581) | 0.79 (1,977) | 0.85 | 0.92 (9,241/820) | 0.87 (1,359) | 0.89 |
| **EqT** | 0.95 (5,385/279) | 0.55 (4,432) | 0.69 | 0.94 (5,942/376) | 0.54 (5,102) | 0.68 |
| | Performance comparison in epicentral distances greater than 200 km | | | | | |
| **OBST** | 0.91 (3,093/309) | 0.76 (953) | 0.83 | 0.89 (3,453/418) | 0.83 (711) | 0.86 |
| **EqT** | 0.89 (329/39) | 0.08 (3,987) | 0.14 | 0.95 (298/16) | 0.07 (4,268) | 0.12 |

Abbreviations: OBST = OBSTransformer, Pre = Precision, Rec = Recall, F1 = F1-score, TP = true positive, FP = false positive, FN = false negative.

**Table 6.** Performance comparison on the AACSE regional-scale test dataset at three different depth ranges.

| | Performance comparison in depths less than 20 km | | | | | |
|---|---|---|---|---|---|---|
| **Model** | **P_Pre (TP/FP)** | **P_Rec (FN)** | **P_F1** | **S_Pre (TP/FP)** | **S_Rec (FN)** | **S_F1** |
| **OBST** | 0.93 (7,593/598) | 0.77 (2,313) | 0.84 | 0.91 (8,872/858) | 0.86 (1,417) | 0.89 |
| **EqT** | 0.93 (4,750/365) | 0.47 (5,389) | 0.62 | 0.94 (5,177/328) | 0.48 (5,642) | 0.63 |
| | Performance comparison in depths greater than 20 km but less than 80 km | | | | | |
| **OBST** | 0.93 (7,082/514) | 0.82 (1,560) | 0.87 | 0.92 (7,748/656) | 0.89 (949) | 0.90 |
| **EqT** | 0.94 (5,007/308) | 0.57 (3,841) | 0.71 | 0.94 (5,082/347) | 0.56 (3,924) | 0.70 |
| | Performance comparison in depths greater than 80 km | | | | | |
| **OBST** | 0.92 (844/78) | 0.77 (247) | 0.84 | 0.94 (949/65) | 0.83 (199) | 0.88 |
| **EqT** | 0.93 (397/31) | 0.35 (741) | 0.51 | 0.97 (432/14) | 0.36 (767) | 0.52 |

Abbreviations: OBST = OBSTransformer, Pre = Precision, Rec = Recall, F1 = F1-score, TP = true positive, FP = false positive, FN = false negative.

**Table 7.** Performance comparison on the Gorda plate local-scale test dataset at three different epicentral distance ranges.

| | Performance comparison in epicentral distances less than 20 km | | | | | |
|---|---|---|---|---|---|---|
| **Model** | **P_Pre (TP/FP)** | **P_Rec (FN)** | **P_F1** | **S_Pre (TP/FP)** | **S_Rec (FN)** | **S_F1** |
| **OBST** | 0.99 (4,764/38) | 0.91 (484) | 0.95 | 0.99 (5,849/61) | 0.95 (302) | 0.97 |
| **EqT** | 0.92 (3,807/332) | 0.77 (1,147) | 0.84 | 0.99 (5,117/49) | 0.83 (1,046) | 0.90 |
| | Performance comparison in epicentral distances greater than 20 km but less than 50 km | | | | | |
| **OBST** | 0.95 (7,924/375) | 0.85 (1,399) | 0.90 | 0.98 (10,841/170) | 0.92 (941) | 0.95 |
| **EqT** | 0.90 (6,358/725) | 0.71 (2,615) | 0.79 | 0.99 (8,837/80) | 0.74 (3,035) | 0.85 |
| | Performance comparison in epicentral distances greater than 50 km | | | | | |
| **OBST** | 0.96 (2,453/105) | 0.82 (553) | 0.88 | 0.95 (3,317/169) | 0.91 (347) | 0.93 |
| **EqT** | 0.90 (1,943/223) | 0.67 (945) | 0.77 | 0.96 (2,425/90) | 0.65 (1,318) | 0.77 |

Abbreviations: OBST = OBSTransformer, Pre = Precision, Rec = Recall, F1 = F1-score, TP = true positive, FP = false positive, FN = false negative.

Table 8. Performance comparison on the Gorda plate local-scale test dataset at three different depth ranges.

| Model | P_Pre (TP/FP) | P_Rec (FN) | P_F1 | S_Pre (TP/FP) | S_Rec (FN) | S_F1 |
|---|---|---|---|---|---|---|
| Performance comparison in depths less than 10 km | | | | | | |
| OBST | 0.95 (1,123/59) | 0.81 (265) | 0.87 | 0.97 (1,520/42) | 0.92 (133) | 0.95 |
| EqT | 0.88 (902/126) | 0.68 (419) | 0.77 | 0.98 (1,265/22) | 0.76 (408) | 0.85 |
| Performance comparison in depths greater than 10 km but less than 20 km | | | | | | |
| OBST | 0.95 (4,403/254) | 0.82 (934) | 0.88 | 0.98 (6,068/146) | 0.92 (547) | 0.94 |
| EqT | 0.88 (3,469/489) | 0.68 (1,633) | 0.76 | 0.98 (4,761/85) | 0.71 (1,915) | 0.83 |
| Performance comparison in depths greater than 20 km | | | | | | |
| OBST | 0.95 (9,599/214) | 0.89 (1,244) | 0.93 | 0.99 (12,344/183) | 0.92 (1,014) | 0.95 |
| EqT | 0.92 (5,996/468) | 0.72 (4,593) | 0.81 | 0.99 (7,604/36) | 0.72 (5,901) | 0.83 |

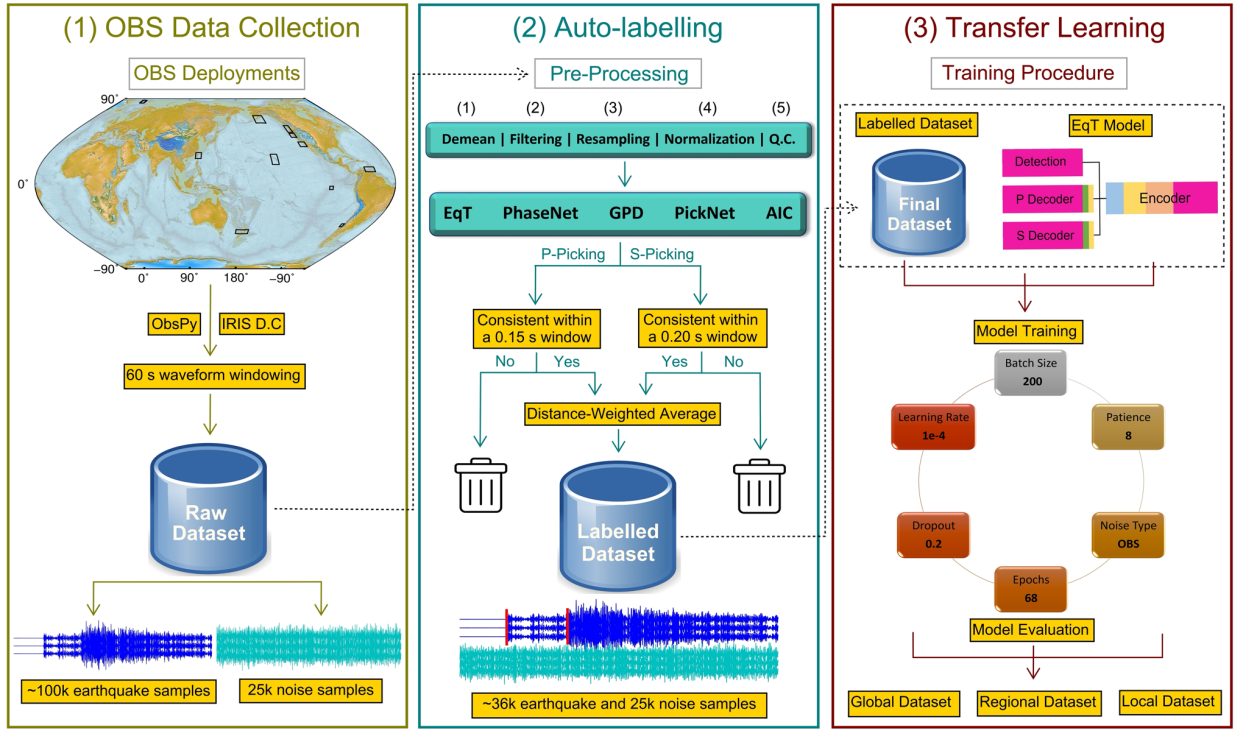

**Figure 1.** Flowchart showing the data collection, training dataset labelling, and transfer learning for the OBSTransformer picker.

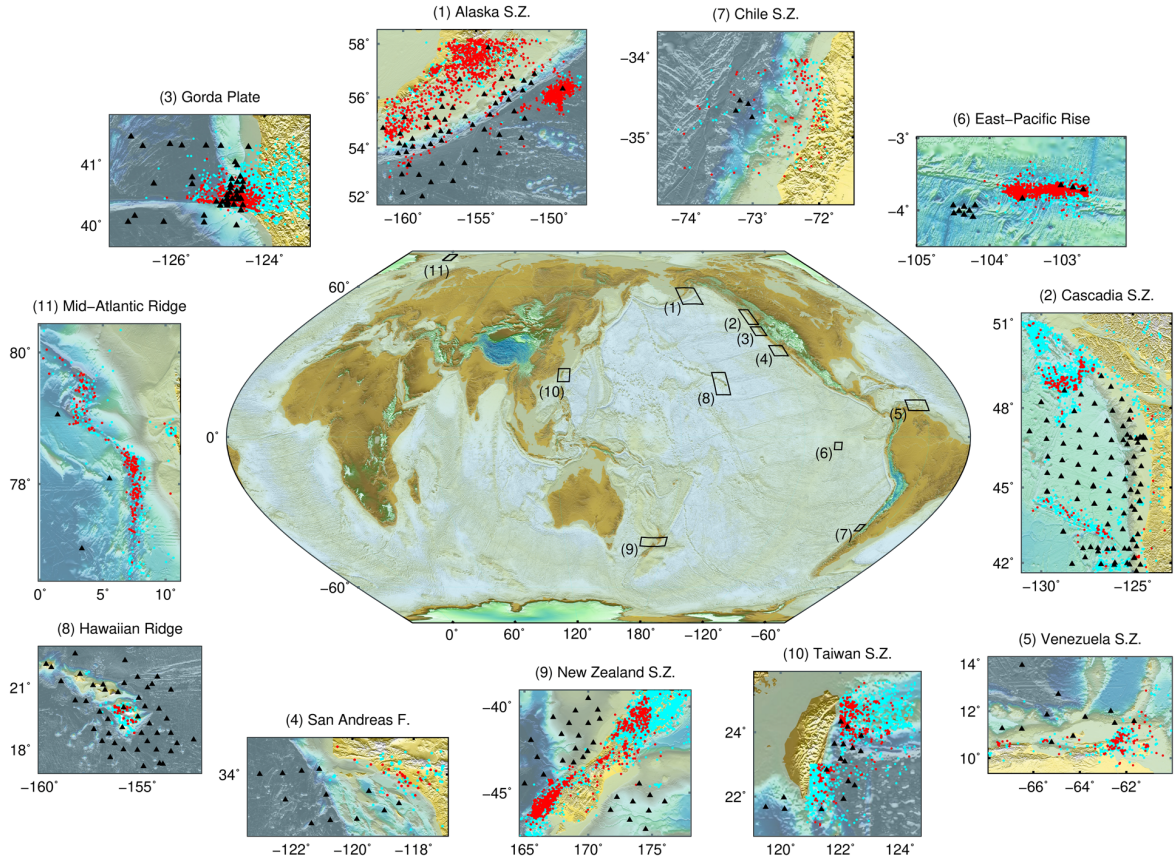

**Figure 2.** Geographical distribution of temporary OBS networks used to compile the training datasets. In each inset, the location of OBS stations (black triangle), earthquakes in our labelled dataset (red dots), and earthquakes discarded during preprocessing (cyan dots) are presented. Earthquake catalogues from Guo et al. (2021) and Gong et al. (2022) are used for the Gorda plate and East Pacific rise, respectively. For other regions, earthquake catalogues are from the IRIS data center.

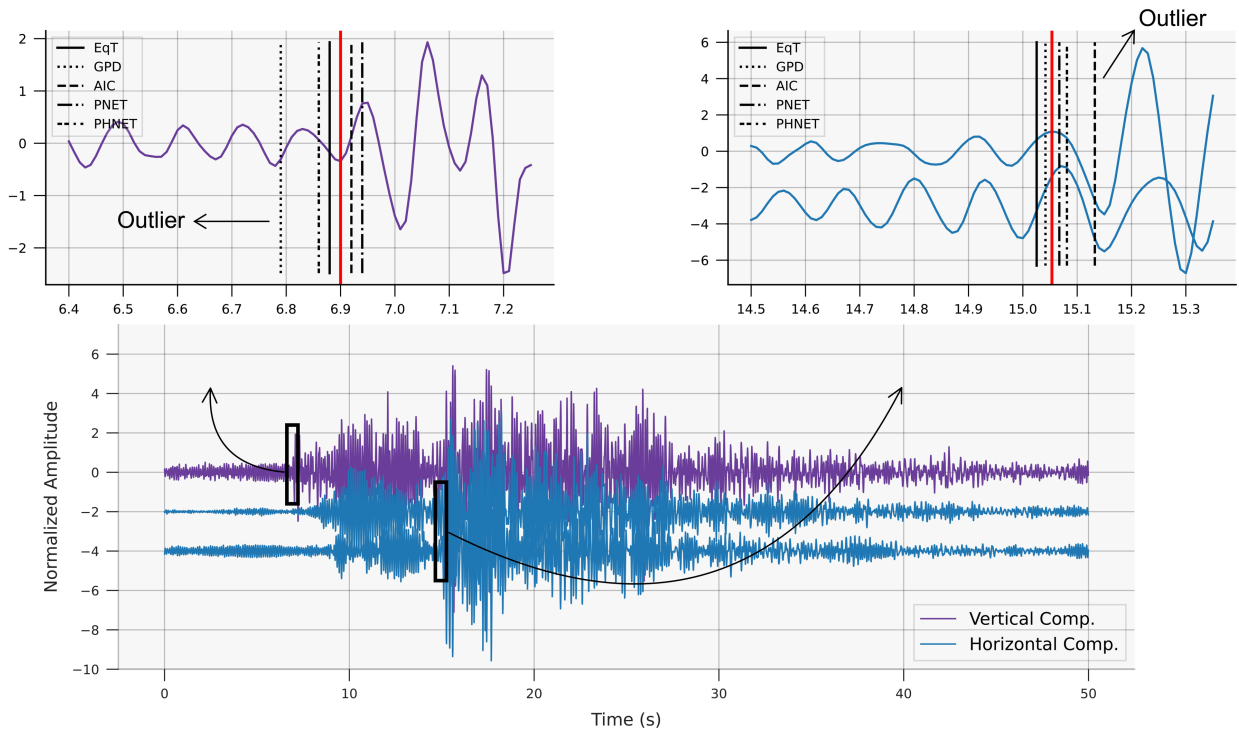

**Figure 3.** An example demonstrates our auto-labelling approach. Picks from different picking methods are marked by different types of lines (see legend). Solid red lines denote the optimal labels. An arrow declares the outliers detected by the skewness test.

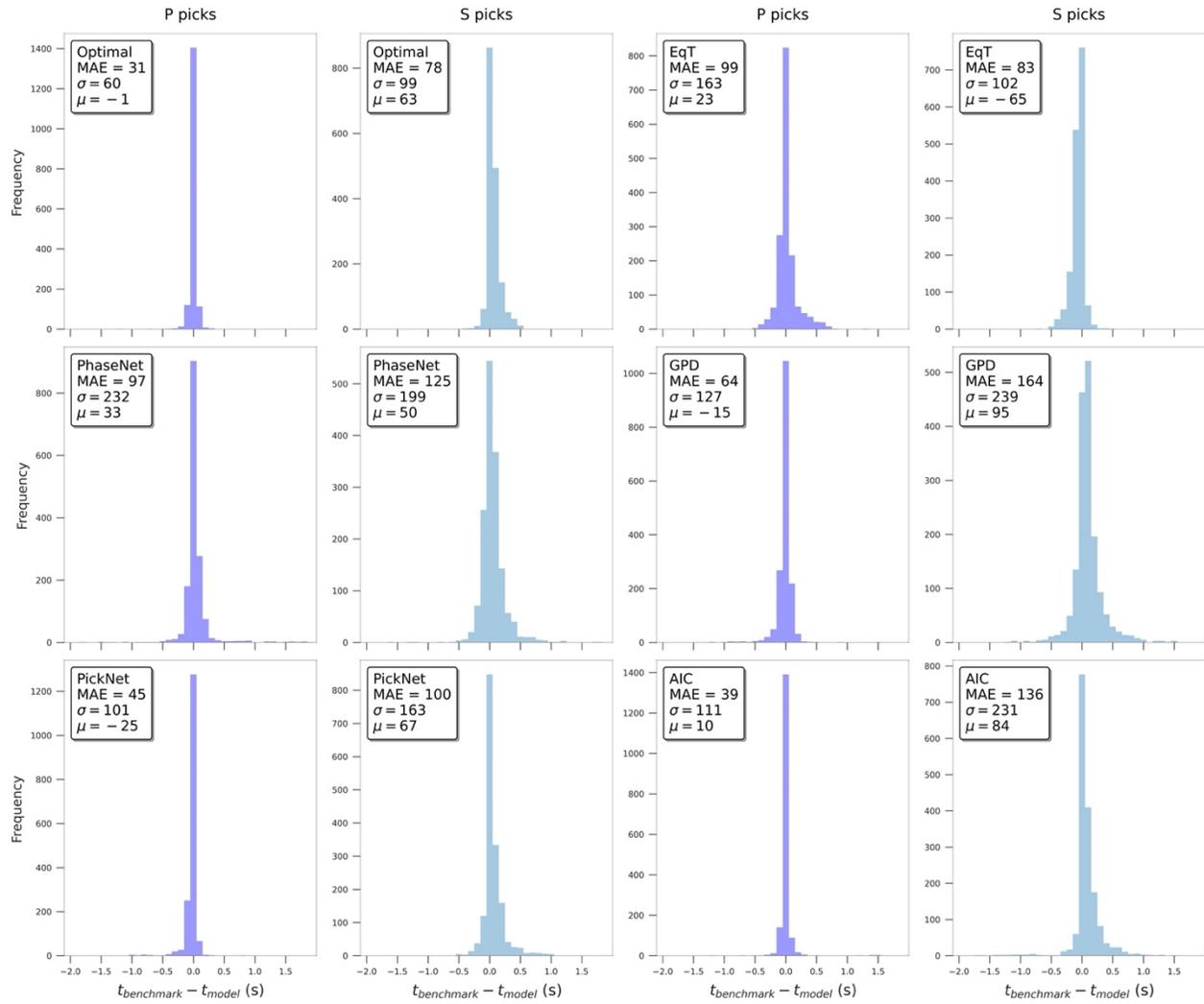

**Figure 4**. Performance comparison among our auto-labelling approach and the individual pickers on the "original" sub-global test dataset. P- and S-wave residual distributions for each picker are shown. MAE, $\sigma$, and $\mu$ correspond to Mean Absolute Error, standard deviation, and mean in milliseconds, respectively.

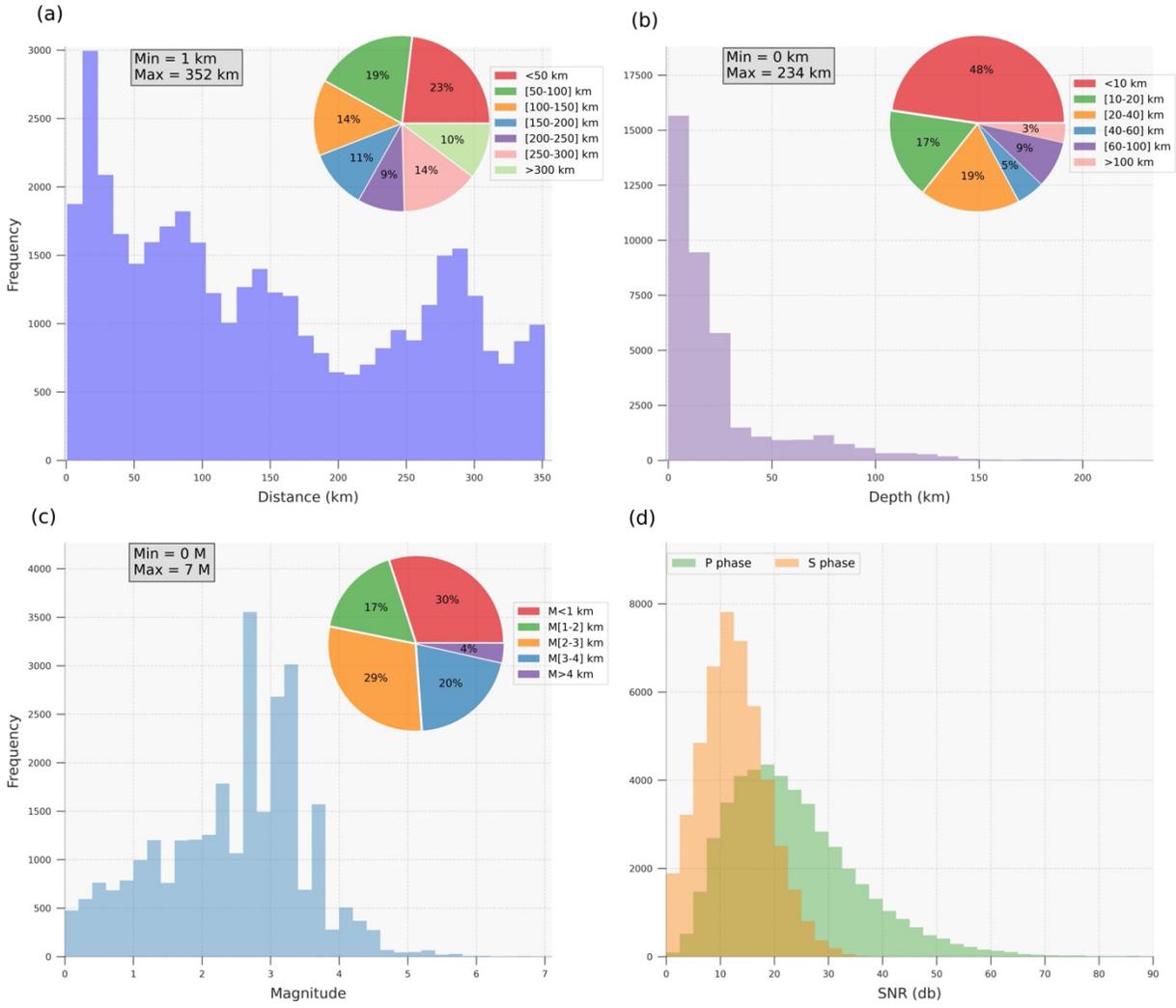

**Figure 5**. The epicentral distance (a), depth (b), magnitude (c), and SNR (d) distributions of training earthquake data as a function of frequency.

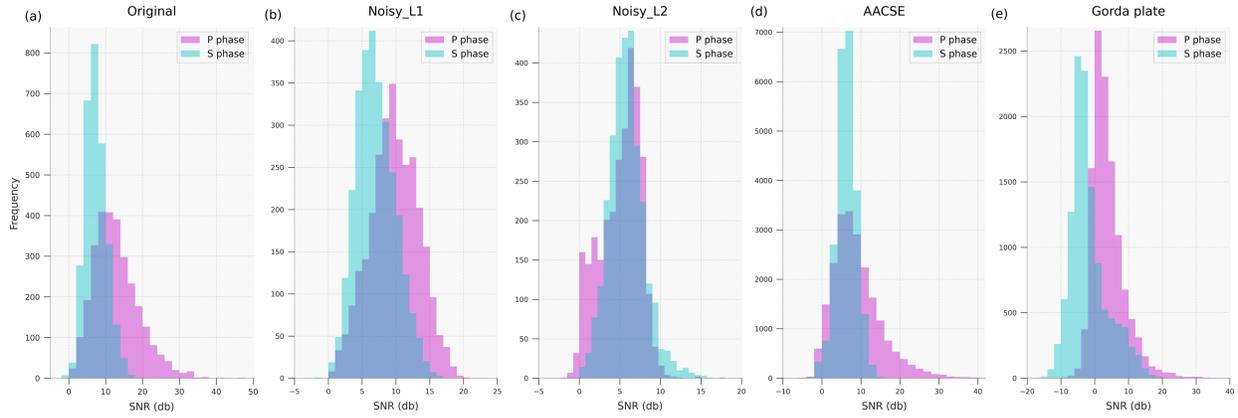

**Figure 6.** SNR distribution of P- and S-phases in the "original" sub-global test dataset (a) and its two noisy variants (b and c), the AACSE test dataset (d), and the Gorda plate test dataset (e). The noisy variants of the "original" test dataset are built by superimposing OBS noise onto the "original" test dataset.

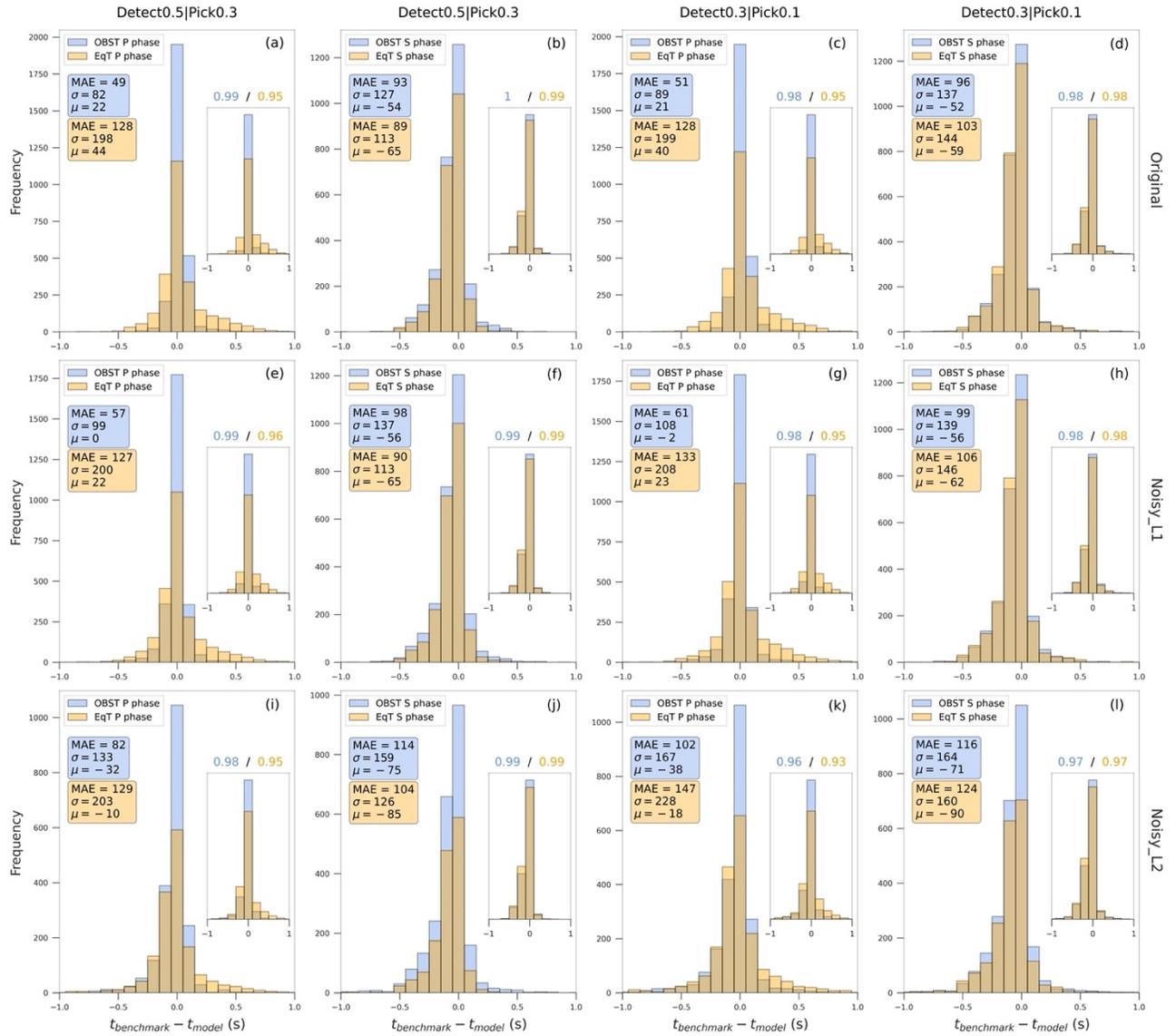

**Figure 7.** Performance comparison between OBST (blue) and EqT (orange) on the "original" (a-d), "Noisy_L1" (e-h), and "Noisy_L2" (i-l) sub-global test datasets. The comparison is made for two detection/picking thresholds of 0.5/0.3 and 0.3/0.1. The insets in each panel display the residual distributions of the common detections by the two pickers, along with the corresponding precisions as headers. The statistics are in milliseconds.

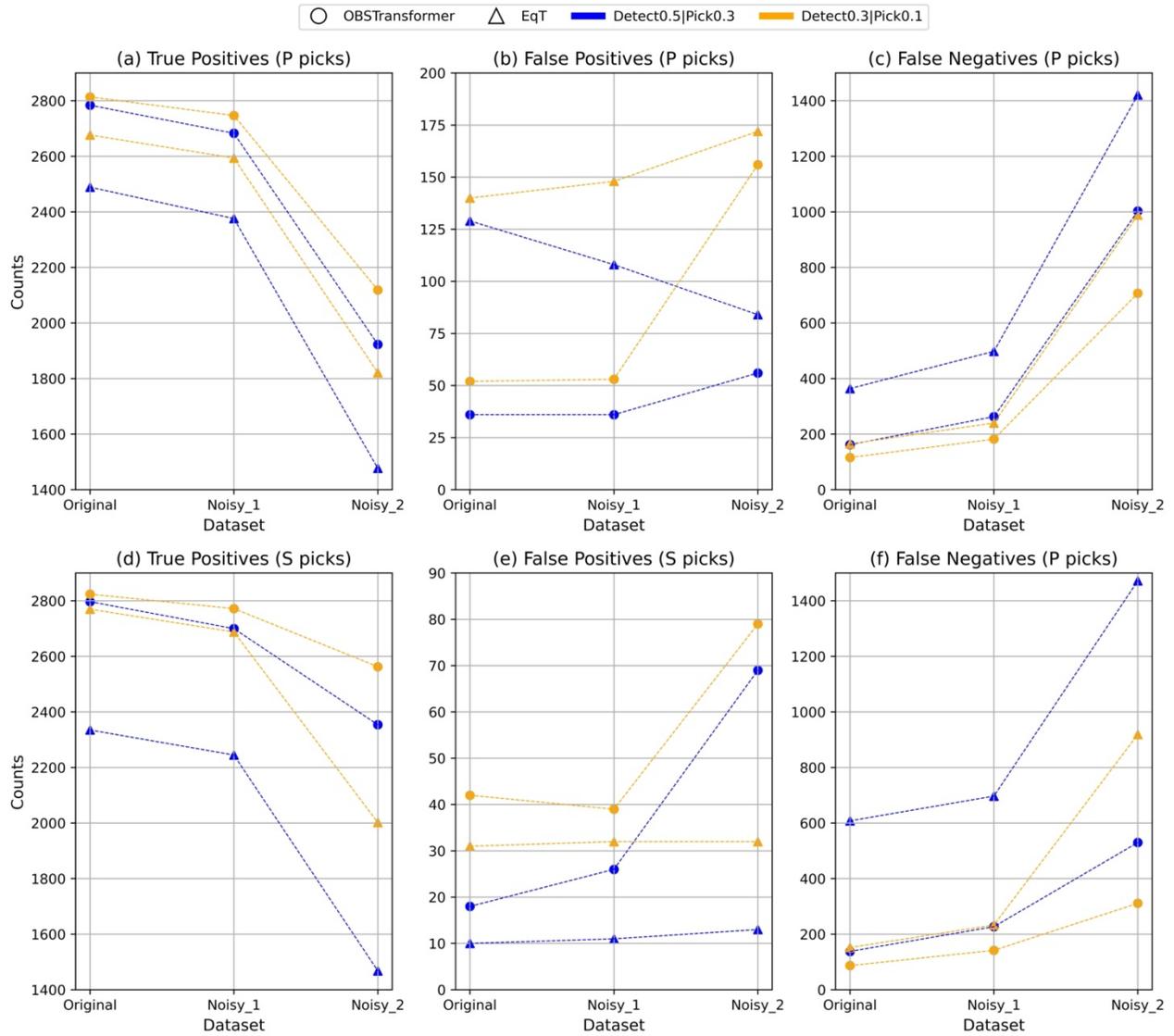

**Figure 8.** Performance comparison in terms of true positives (a, d), false positives (b, e), and false negatives (c, f) for P (top) and S (bottom) phases. The blue and orange circles (OBST) and triangles (EqT) correspond to the detection/picking thresholds of 0.5/0.3 and 0.3/0.1, respectively. In each figure, the comparison is made across three different noise levels, progressively increasing from left to right.

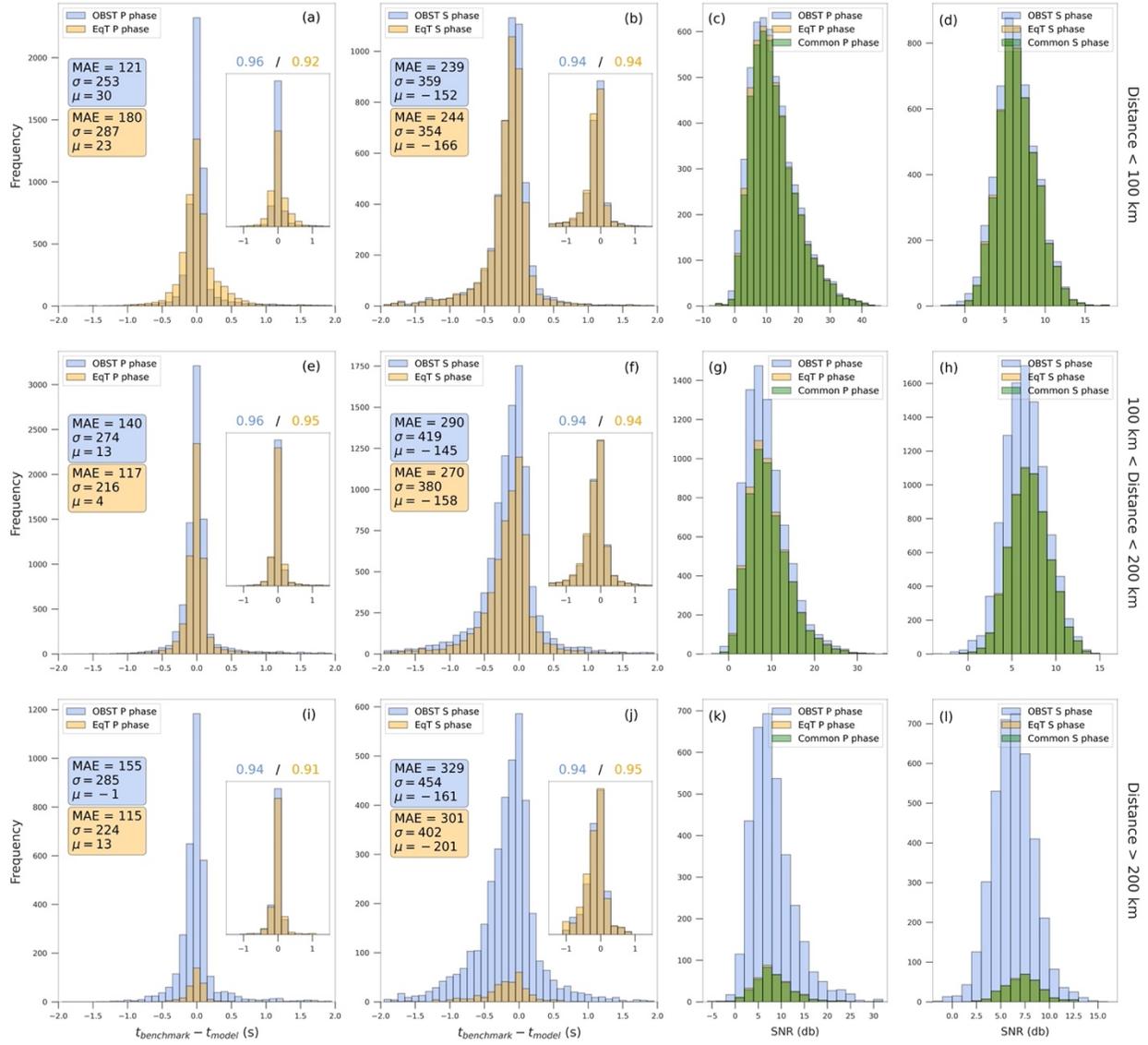

**Figure 9.** Residual distributions of P and S phases by OBST (blue) and EqT (orange) on the AACSE test dataset at three different epicentral distance ranges: (a-d) less than 100 km, (e-h) larger than 100 km but less than 200 km, and (i-l) larger than 200 km. MAE, $\sigma$, and $\mu$ are the Mean Absolute Error, standard deviation, and mean of residuals in milliseconds, respectively. SNR distributions of detected P and S phases are presented on the right half of the figure. The green bars correspond to the common detections.

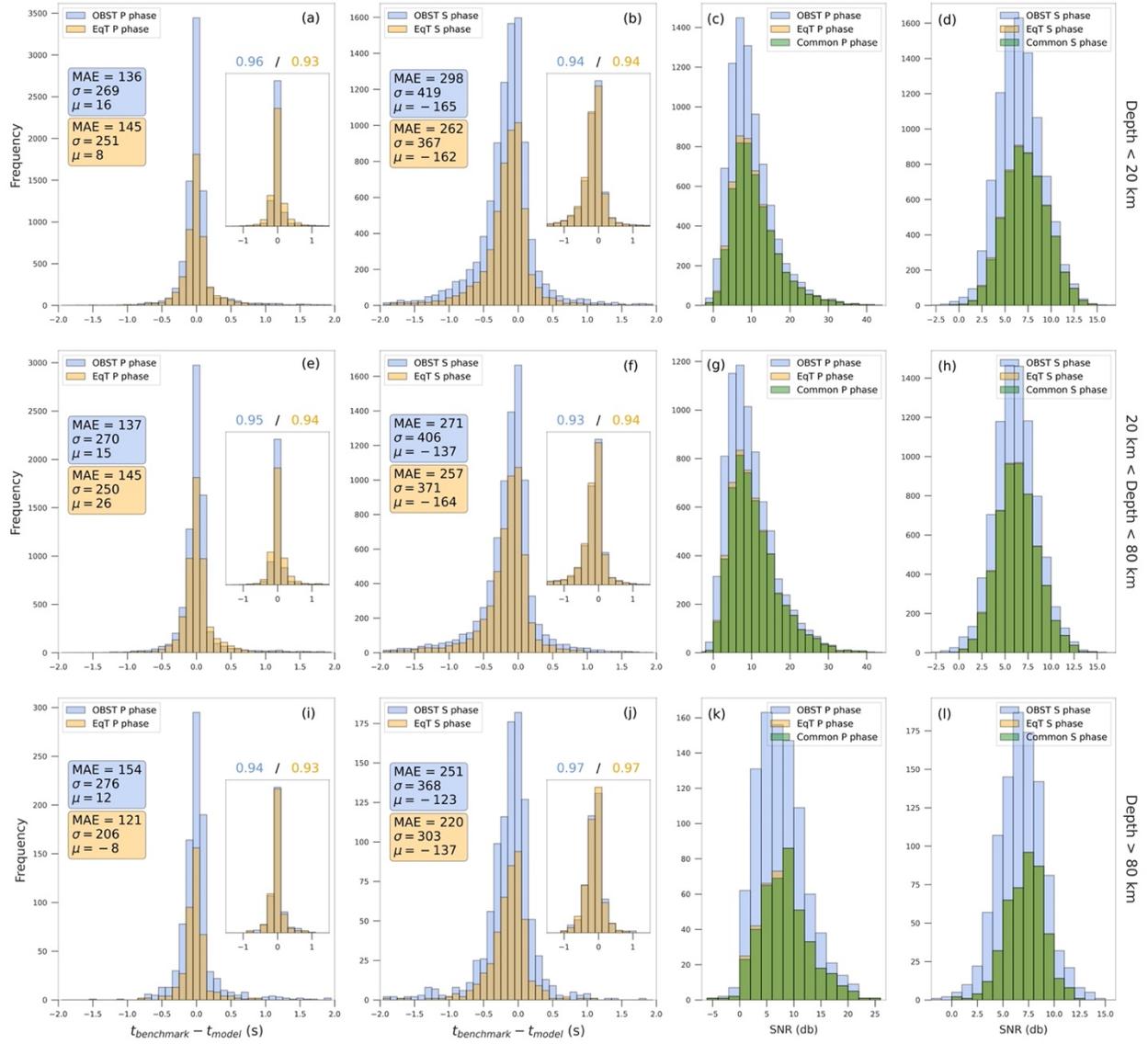

**Figure 10.** Residual distributions of P and S phases by OBST (blue) and EqT (orange) on the AACSE test dataset at three different depth ranges: (a-d) shallower than 20 km, (e-h) deeper than 20 km but shallower than 80 km, and (i-l) deeper than 80 km. MAE, $\sigma$, and $\mu$ are the Mean Absolute Error, standard deviation, and mean of residuals in milliseconds, respectively. SNR distributions of detected P and S phases are presented on the right half of the figure. The green bars correspond to the common detections.

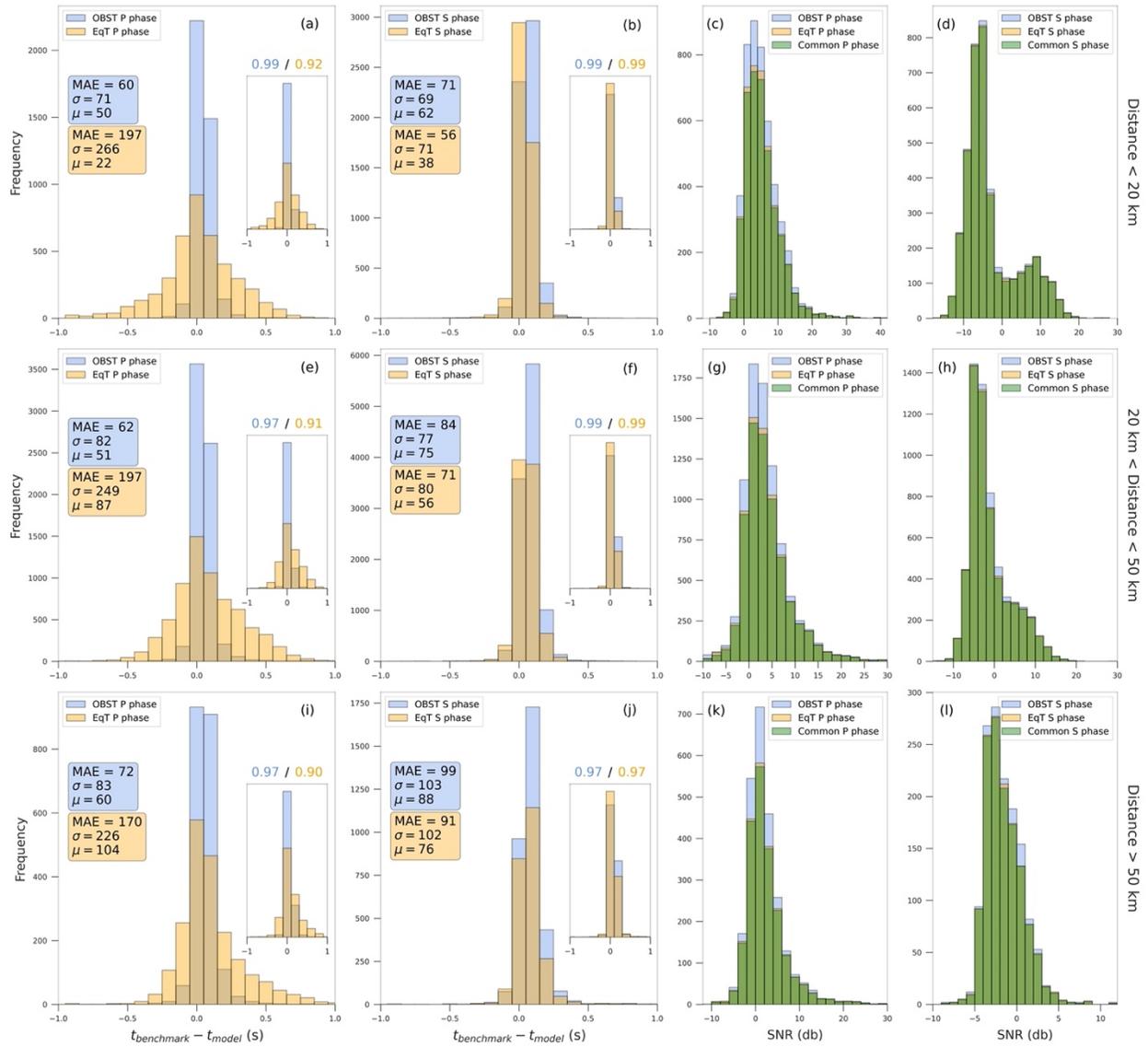

**Figure 11.** Residual distributions of P and S phases by OBST (blue) and EqT (orange) on the Gorda plate test dataset at three different epicentral distance ranges: (a-d) less than 20 km, (e-h) larger than 20 km but less than 50 km, and (i-l) larger than 50 km. MAE, σ, and μ are the Mean Absolute Error, standard deviation, and mean of residuals in milliseconds, respectively. SNR distributions of detected P and S phases are presented on the right half of the figure. The green bars correspond to the common detections.

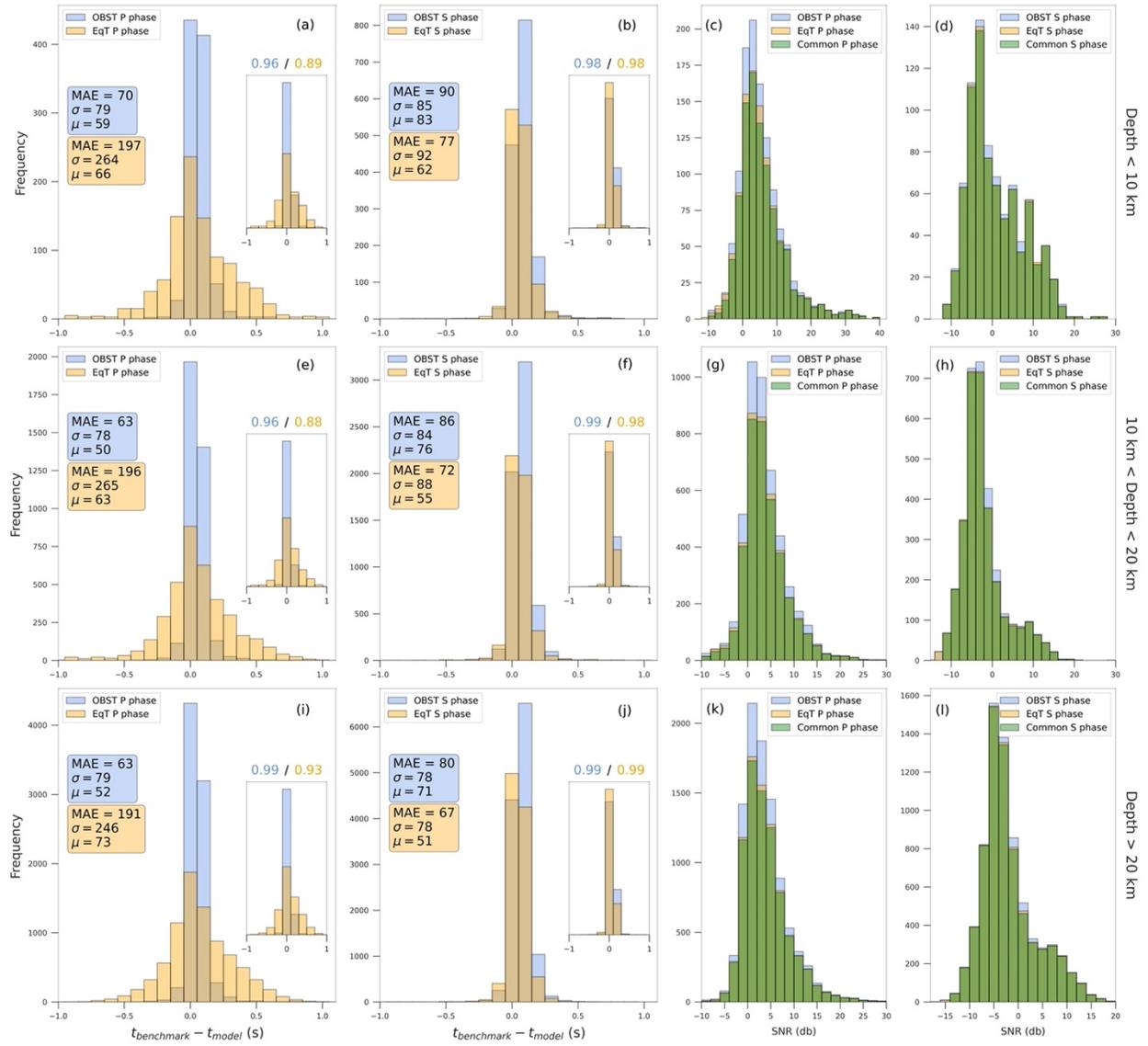

**Figure 12.** Residual distributions of P and S phases by OBST (blue) and EqT (orange) on the Gorda plate test dataset at three different depth ranges: (a-d) shallower than 10 km, (e-h) deeper than 10 km but shallower than 20 km, and (i-l) deeper than 20 km. MAE, σ, and μ are the Mean Absolute Error, standard deviation, and mean of residuals in milliseconds, respectively. SNR distributions of detected P and S phases are presented on the right half of the figure. The green bars correspond to the common detections.

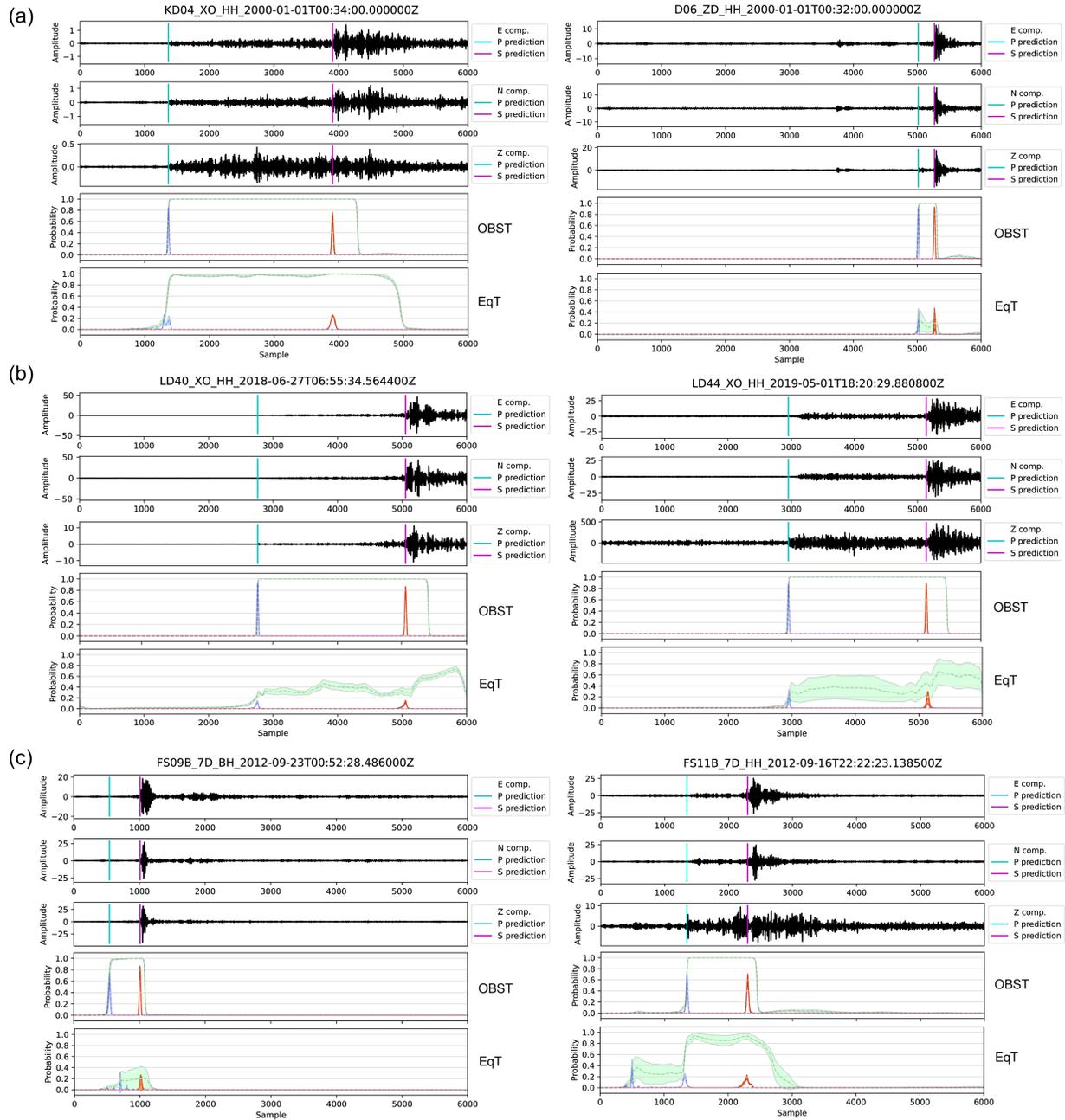

**Figure 13**. Examples of phase picking by OBST and EqT from the sub-global (a), regional-scale (b), and local-scale (c) tests. Benchmarks are marked by the solid cyan (P) and pink (S) vertical lines on seismograms. The lower panels of seismograms show blue, red, and green distributions, which correspond to P, S, and detection probabilities, respectively. Shades surrounding each probability distribution with the same colour represent the associated uncertainties. The headers contain trace information.

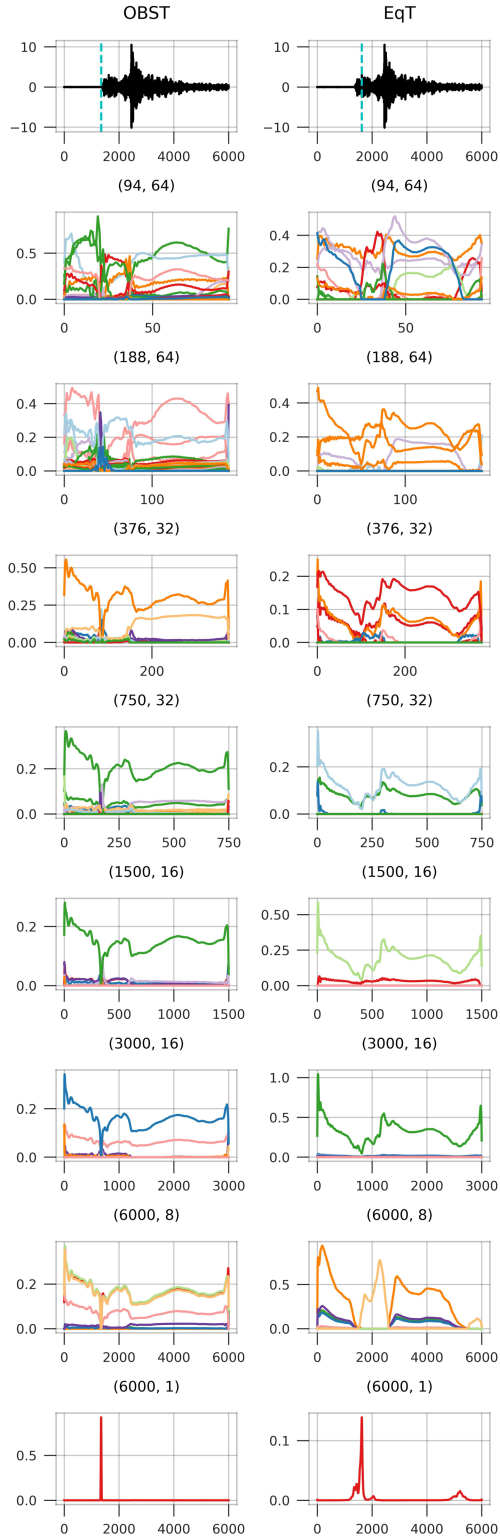

**Figure 14**. Feature maps of the CNN layers in the P-decoder of OBST (left) and EqT (right) for a sample input. The vertical component of the input seismogram is shown on the top, followed by the CNN layers (Conv1D_31 to Conv1D_37) of the decoder. The decoder output is shown on the bottom. The headers in each panel indicate the size (left) and number (right) of filters in each layer. The corresponding arrivals by the two pickers are marked on the waveforms (cyan dashed lines).